\documentclass{article}
\pdfoutput=1
\usepackage{graphicx}  
\usepackage{amsmath}   
\usepackage[compress]{cite}
\usepackage{amssymb}   
\usepackage{bm} 
\usepackage{dcolumn}
\usepackage{color}
\usepackage{mathrsfs}
\usepackage{amsfonts}
\usepackage{varioref}
\usepackage{float}
\restylefloat{table}
\usepackage[caption=false]{subfig}
\RequirePackage[colorlinks,citecolor=blue,urlcolor=magenta,linkcolor=blue]{hyperref}
\input epsf
\usepackage[labelsep=none,labelformat=empty]{caption}
\def\gr{general relativity}
\def\RN{Reissner-Nordstr\"{o}m }
\def\KN{Kerr-Newmann }

\labelformat{section}{Section #1} 
\labelformat{subsection}{Section #1} 
\labelformat{subsubsection}{Section #1}
\labelformat{subsubsubsection}{Section #1}
\labelformat{equation}{Eq.~(#1)} 
\labelformat{figure}{Fig.~#1} 
\labelformat{subfigure}{Fig.~\thefigure#1} 
\labelformat{table}{Table.~#1} 
\labelformat{appendix}{Appendix #1}

\title{Does black hole continuum spectrum signal $f(R)$ gravity in higher dimensions?}
\author{Indrani Banerjee\footnote{tpib@iacs.res.in}~$^{1}$, Bhaswati Mandal\footnote{tpbm3@iacs.res.in}~$^{1}$ and Soumitra SenGupta\footnote{tpssg@iacs.res.in}~$^{1}$\\
{\small{$^{1}$School of Physical Sciences, Indian Association for the Cultivation of Science, Kolkata-700032, India}}}

\date{}

\begin{document}

\maketitle

\begin{abstract}
Extra dimensions, which led to the foundation and inception of string theory, provide an elegant approach to force-unification. With bulk curvature as high as the Planck scale, higher curvature terms, namely $f(R)$ gravity seems to be a natural addendum in the bulk action. 
These can not only pass the classic tests of \gr\ but also serve as potential alternatives to dark matter and dark energy. With interesting implications in inflationary cosmology, gravitational waves and particle phenomenology it is worth exploring the impact of extra dimensions and $f(R)$ gravity in black hole accretion. Various classes of black hole solutions have been derived which bear non-trivial imprints of these ultraviolet corrections to general relativity. This in turn gets engraved in the continuum spectrum emitted by the accretion disk around black holes. Since the near horizon regime of supermassive black holes manifest maximum curvature effects, we compare the theoretical estimates of disk luminosity with quasar optical data to discern the effect of the modified background on the spectrum. In particular, we explore a certain class of black hole solution bearing a striking resemblance with the well-known \RN de Sitter/anti-de Sitter/flat spacetime which unlike general relativity can also accommodate a negative charge parameter. By computing error estimators like chi-square, Nash-Sutcliffe efficiency, index of agreement, etc. we infer that optical observations of quasars favor a \emph{negative} charge parameter which can be a possible indicator of extra dimensions. The analysis also supports an asymptotically de Sitter spacetime with an estimate of the magnitude of the cosmological constant whose origin is solely attributed to $f(R)$ gravity in higher dimensions.

\end{abstract}

\section{INTRODUCTION}
General relativity (GR) is a classic example of a scientific theory that is elegant, simple and powerful. Till date, it is the most successful theory of gravity in explaining a plethora of observations namely, the perihelion precession of mercury, the bending of light, the gravitational redshift of radiation from distant stars, to name a few \cite{Will:2005yc,Will:2005va,Yunes:2013dva}. Very recently, the shadow of the black hole in M87 observed by the Event Horizon Telescope has further added to its phenomenal success \cite{Akiyama:2019cqa,Akiyama:2019eap,Ayzenberg:2018jip}. Yet it is instructive to subject GR to further tests since it is marred with unresolved issues like singularities \cite{Penrose:1964wq,Hawking:1976ra,Christodoulou:1991yfa} and falls short in explaining the nature of dark energy and dark matter \cite{Milgrom:1983pn,Bekenstein:1984tv,Milgrom:2003ui,Perlmutter:1998np,Riess:1998cb}. Moreover, the quantum nature of gravity is still elusive and ill-understood \cite{Rovelli:1996dv,Dowker:2005tz,Ashtekar:2006rx}. All this makes the quest for a more complete theory of gravity increasingly compelling such that it yields GR in the low energy limit. Consequently a surfeit of alternate gravity models are proposed which can potentially fulfill the deficiencies in GR. A viable alternate gravity theory must be free from ghost modes, be consistent with solar system based tests, should not engender a fifth force in local physics and should successfully explain observations that GR fails to address. The alternate gravity models which fulfill these benchmark can be broadly classified into three categories: (i) Modified gravity models where the gravity action is supplemented with higher curvature terms, e.g., f(R) gravity \cite{Nojiri:2010wj,Nojiri:2006gh,Capozziello:2006dj,Bahamonde:2016wmz}, Lanczos-Lovelock models etc. \cite{Lanczos:1932zz,Lanczos:1938sf,Lovelock:1971yv,Padmanabhan:2013xyr,Dadhich:2015ivt}
(ii) Extra-dimensional models that alter the effective 4-dimensional gravitational field equations due to the bulk Weyl stresses and higher order corrections to the stress-tensor \cite{Shiromizu:1999wj,Dadhich:2000am,Harko:2004ui,
Carames:2012gr,Haghani:2012zq,Chakraborty:2014xla,Chakraborty:2015bja} and (iii) Scalar-tensor theories of gravity which include the Brans-Dicke theory and the more general Horndeski models \cite{Brans:1961sx,Horndeski:1974wa,Sotiriou:2013qea,Babichev:2016rlq}.

In this work we will consider modifications to the gravity sector by introducing $f(R)$ gravity in five dimensions. Among the various modified gravity models, $f(R)$ theories have attracted the attention of physicists for a long time \cite{Nojiri:2010wj,Sotiriou:2008rp,DeFelice:2010aj,Chakraborty2015} since they invoke the simplest modification to the Einstein-Hilbert action and yet exhibit sufficient potential to address a host of cosmological and astrophysical observations. These include, but are not limited to, the late time acceleration \cite{Bamba:2008ja,delaCruzDombriz:2006fj,Nojiri:2017ncd} and the initial power-law inflation of the universe \cite{Starobinsky:1980te,Nojiri:2017ncd}, the four cosmological phases \cite{Nojiri:2006gh,Nojiri:2008fk}, the rotation curves of spiral galaxies \cite{Capozziello:2006uv,Capozziello:2006ph} and the detection of gravitational waves \cite{Corda:2009re,Corda:2009kc,Corda:2008si}. Although these models are plagued with ghost modes, certain $f(R)$ models e.g. $f(R)$ theory on a constant curvature hypersurface can be shown to be ghost free \cite{Nojiri:2007jr,Nojiri:2007as,Nojiri:2007cq}. In addition, they can successfully surpass the solar system tests which only impose constraints on $f^{\prime\prime}(R)$ and hence on the model parameters \cite{Capozziello:2005bu,Capozziello:2007ms,Capozziello:2006jj}.

Extra-dimensions on the other hand were mainly invoked to provide a framework to unify gravity and electromagnetism \cite{Nordstrom:1988fi,Kaluza:1921tu,Klein:1926fj}. This subsequently provided a framework for string theory and M-theory that succeeded in unifying all the known forces under a single umbrella \cite{Horava:1995qa,Polchinski:1998rq,Polchinski:1998rr}. 
The large radiative corrections to the Higgs mass arising due to the huge disparity between the electro-weak scale and the Planck scale \cite{Antoniadis:1990ew,ArkaniHamed:1998rs,Antoniadis:1998ig,Randall:1999ee,Csaki:1999mp} led to the emergence of a diversity of string inspired brane-world models.   
Most of these models assume that the observable universe is confined in a 3-brane where all 
the Standard Model particles and fields reside while gravity permeates to the bulk \cite{Antoniadis:1990ew,Antoniadis:1998ig,ArkaniHamed:1998rs,Randall:1999vf,Garriga:1999yh,Randall:1999ee,Csaki:1999mp}.
They possess interesting phenomenological implications \cite{ArkaniHamed:1998nn, Davoudiasl:1999tf, Davoudiasl:2000wi, Davoudiasl:1999jd,Hundi:2011dc, Chakraborty:2014zya} and distinct observational signatures including production of mini-black holes which can be tested in present and future collider experiments \cite{Dimopoulos:2001hw, Banks:1999gd}. In the galactic scale, they offer an alternative to the elusive dark matter \cite{Pal:2004ii,Capozziello:2006uv,Boehmer:2007az,Harko:2007yq,Chakraborty:2015zxc} while in cosmology they have interesting implications in the inflationary epoch \cite{Lukas:1999yn,ArkaniHamed:1999gq,Dienes:1998hx,Mazumdar:2000sw,Chakraborty:2013ipa,Banerjee:2017lxi,Banerjee:2018kcz} and also serve as a possible proxy to dark-energy \cite{Koyama:2003be,Mazumdar:2000gj,Maartens:2000fg,Maartens:2003tw,Haghani:2012zq,Binetruy:1999ut,Csaki:1999jh}. Since the ultraviolet nature of gravity is unknown, it is often believed that in the high energy regime, the deviations from Einstein gravity may manifest through the existence of extra dimensions. Moreover, the bulk curvature is expected to be as high as the Planck scale and hence higher order corrections to the gravity action should become relevant in the high energy regime. 

In this work we consider a single braneworld scenario with a positive tension which is embedded in a five dimensional bulk containing $f(R)$ gravity. 
The addition of $f(R)$ in higher dimensions cause substantial modification to the effective gravitational field equations on the brane \cite{Bazeia:2013oha,Carames:2012gr,Borzou:2009gn, Shiromizu:1999wj, Chakraborty:2015bja} which are obtained from Gauss-Codazzi equation and the junction conditions \cite{gravitation}. Such deviations from Einstein's equations are expected to become more conspicuous in the high energy/high curvature domain. Therefore, the near horizon regime of black holes where the curvature effects are maximum, seem to be an ideal astrophysical laboratory to test these models against observations.

Various classes of vacuum solutions of these field equations have been obtained \cite{Dadhich:2000am,Harko:2004ui,Chakraborty:2014xla, Germani:2001du,Aliev:2005bi} which possess distinct signatures of extra dimensions and $f(R)$ gravity. In the event the vacuum solutions are static and spherically symmetric, the electric part of the Weyl tensor can be decomposed into terms involving ``dark radiation" and ``dark pressure". Suitable integrability conditions lead to different classes of vacuum solutions which determine the spacetime geometry. The solutions thus derived exhibit substantial modification from the well-known Schwarzschild spacetime which are attributed to the non-local effects of the bulk Weyl tensor and $f(R)$ gravity in the action. These deviations in the background spacetime are sculpted in the continuum spectrum emitted from the accretion disk around black holes. In particular, since the curvature effects are maximum in supermassive black holes, the quasar continuum spectra can act as potential astrophysical probes to establish/falsify/constrain these models.

In a recent work \cite{Banerjee:2017hzw} we explored an exact black hole solution in the brane with bulk Einstein gravity. It resembles the well-known \RN spacetime in \gr$~$where the tidal charge parameter can assume both signatures. By comparing the disk luminosity of quasars in such a background with the corresponding observations we conclude that a negative charge parameter is favored which is characteristic to braneworld black holes. Adding $f(R)$ gravity in the bulk action adds a vaccum energy term to the aforesaid black hole solution where the cosmological constant owes its origin to terms involving $f(R)$ gravity in higher dimensions. 
In this work we investigate the effect of such a spacetime on the quasar continuum spectrum which enables us to explore the signature of the tidal charge parameter in the presence of the cosmological constant term in the metric. Subsequently,  
we also derive constraints on the magnitude of the cosmological constant from quasar optical data. Further, we also investigate the effect of other black hole solutions on the quasar continuum spectrum, which are derived by altering the relations connecting the ``dark radiation" and ``dark pressure". 

The paper is organized as follows: In \ref{S2} we discuss the modifications induced in the gravitational field equations due to the presence of bulk $f(R)$ gravity. The static, spherically symmetric, vaccum solutions of these field equations are reviewed in \ref{S3}. In \ref{S4} we examine the properties of the black hole continuum spectrum in presence of the background spacetimes discussed in \ref{S3}. \ref{S5} is dedicated to  numerical analysis where the theoretically computed luminosities from the accretion disk of eighty quasars are compared with the corresponding observed values. Finally, we conclude with a summary and the discussion of our results in \ref{S6}.

\textit{Notations and Conventions:} Throughout this paper, the Greek indices denote the four dimensional spacetime and capitalized latin alphabets represent the five dimensional bulk indices. We will work in geometrized unit with $G=1=c$ and the metric convention will be mostly positive.

\section{Static, spherically symmetric black hole solutions in higher dimensional $f(\mathcal{R})$ gravity}\label{S2}
In this section we consider $f(R)$ gravity in the bulk action and derive the effective gravitational field equations on the brane. The bulk action 
$\mathcal{A}$ assumes the form,
\begin{equation}
\mathcal{A} = \int d^{5}x\sqrt{-G}\left[\frac{f(R)}{2\kappa_{5}^{2}}+\mathcal{L}_{m}\right]
\end{equation}
where  $G_{AB}$ is the bulk metric, $R$ is the bulk Ricci scalar and $\mathcal{L}_{m}$ is the matter Lagrangian. The bulk indices are denoted by capitalized latin alphabets e.g. A, B which run over all space-time dimensions while Greek letters denote the brane coordinates. The gravitational field equation obtained by varying the bulk action with respect to $G_{AB}$ is given by,
\begin{equation}
f^{\prime}(R)R_{AB}-\frac{1}{2}G_{AB}f(R)+G_{AB}\Box f^{\prime}(R)-\nabla_{A}\nabla_{B}f^{\prime}(R)=\kappa^{2}_{5}T_{AB}
\end{equation}
where $R_{AB}$ is the bulk Ricci tensor, $\kappa^{2}_{5} = 8\pi G_{5}$ is the five dimensional gravitational constant and prime denotes derivative with respect to $R$.
The bulk energy-momentum tensor can be written as,
\begin{align}
T_{AB} = - \Lambda_{5}G_{AB} + \delta(\phi) (- \lambda_{T}g_{\mu\nu} + \tau_{\mu\nu}) e_{A}^{\mu} e_{B}^{\nu}
\end{align}
where $\Lambda_{5}$, the negative vacuum energy density on the bulk, the brane tension $\lambda_T$ and the brane energy-momentum tensor $\tau_{\mu\nu}$ are the sources of the gravitational field on the bulk.  The various physical quantities on the bulk are projected onto the brane with the help of the projector $e^{\mu}_{A}$. The brane is located at $\phi=0$ (where $\phi$ represents the extra coordinate) and the induced metric on the $\phi=0$ hypersurface is represented by $g_{\mu\nu}$.

In order to obtain the effective gravitational field equations on the brane, Gauss-Codazzi equation is used which connects the bulk Riemann tensor to that of the brane with the help of the projector $e^{\mu}_{A}$ and the extrinsic curvature tensor $K_{\mu\nu}$. The extrinsic curvature is related to the covariant derivative of the normalized normals to the brane $n^A$ and encodes the embedding of the brane into the bulk. The presence of a brane energy momentum tensor leads to a discontinuity in $K_{\mu\nu}$ across the brane. Israel junction conditions and a $Z_2$ orbifold symmetry relates this discontinuity in the extrinsic curvature to the brane energy momentum tensor. For a detailed derivation one is referred to \cite{gravitation,Harko:2004ui,Chakraborty:2014xla,Chakraborty:2015bja}. 

With the above considerations the effective four-dimensional gravitational field equations on the brane assume the form,
\begin{align} 
\mathcal{R}_{\mu\nu} - \frac{1}{2}\mathcal{R}g_{\mu\nu} = -\Lambda_{4}g_{\mu\nu} + 8\pi G_{4}\tau_{\mu\nu} + \kappa^{4}_{5}\pi_{\mu\nu}+ Q_{\mu\nu}- E_{\mu\nu}  \label{EE}
\end{align} 
where
\begin{align}
\Lambda_{4} &\boldsymbol= \frac{1}{2}\kappa^{2}_{5}\bigg{[}\frac{\Lambda_{5}}{f^{\prime}(R)}+\frac{1}{6}\kappa^{2}_{5}\lambda^{2}_{T}\bigg{]} \label{5}\\
G_{4} &= \frac{\kappa^{4}_{5}\lambda_{T}}{48\pi} \label{6}\\
\pi_{\mu\nu} &=-\frac{1}{4}\tau_{\mu\alpha}\tau^{\alpha}_{\nu} + \frac{1}{12}\tau\tau_{\mu\nu} + \frac{1}{8}g_{\mu\nu}\tau_{\alpha\beta}\tau^{\alpha\beta} - \frac{1}{24}g_{\mu\nu}\tau^{2} \label{7}\\
Q_{\mu\nu} &= \bigg{[}h(R)g_{\mu\nu} + \frac{2}{3}\frac{\nabla_{A}\nabla_{B}f^{\prime}(R)}{f^{\prime}(R)}(e^{A}_{\mu}e^{B}_{\nu} + n^{A}n^{B}g_{\mu\nu})\bigg{]}_{\phi=0} \label{8}\\
h(R) &= \frac{1}{4}\frac{f(R)}{f^{\prime}(R)} - \frac{1}{4}R - \frac{2}{3}\frac{\Box f^{\prime}(R)}{f^{\prime}(R)} \\
E_{\mu\nu} &=C_{ABCD}e^{A}_{\mu}n^{B}e^{C}_{\nu}n^{D} \label{9}\\
\end{align}

In \ref{EE}, $\mathcal{R}_{\mu\nu}$ and $\mathcal{R}$ refer to the Ricci tensor and Ricci scalar on the brane while $\Lambda_{4}$ and $G_{4}$ represent the 4-dimensional cosmological constant and gravitational constant respectively. \ref{5} serves as the fine balancing relation of the Randall-Sundrum single brane model \cite{Randall:1999vf, Shiromizu:1999wj} which enables the brane tension to be tuned appropriately with the bulk cosmological constant to yield de-Sitter, anti de-Sitter or flat branes.  
In \ref{EE}, $\pi_{\mu\nu}$ represents higher order terms associated with the brane energy momentum tensor due to the local effects of the bulk on the brane. The term $Q_{\mu\nu}$ arises because of the presence of higher curvature terms in the bulk action. In the event $f(R) = R$, $Q_{\mu\nu}=0$ and we recover the projected field equations on the brane due to pure Einstein gravity in the bulk. The expression for $Q_{\mu\nu}$ can be simplified further by assuming that $\partial_\mu R=0$ when the second term in \ref{8} vanishes (see for example \cite{Chakraborty:2014xla}) such that, 
\begin{align}
Q_{\mu\nu}&= \bigg{[}h(R)g_{\mu\nu} + \frac{2}{3}\frac{\nabla_{A}\nabla_{B}f^{\prime}(R)}{f^{\prime}(R)}n^{A}n^{B}g_{\mu\nu}\bigg{]}_{\phi=0}=\mathcal{F}(R)g_{\mu\nu}\label{10}
\end{align}
Since the bulk Ricci scalar is expected to be a well-behaved quantity, it can be expanded in a Taylor series around $\phi=0$, i.e., 
\begin{align}
R=R_0+R_1 \phi + R_2\frac{\phi^2}{2} +\mathcal{O}(\phi^3) \label{11}
\end{align}
where the coefficients are constants since $R$ is independent of the brane coordinates.
This implies that the derivatives of $R$ evaluated at $\phi=0$ in \ref{10} will result in a constant contribution independent of the brane coordinates.

The last term on the right hand side of \ref{EE} is $E_{\mu\nu}$ which epitomizes the electric part of the bulk Weyl tensor with its origin in the nonlocal effect from the free bulk gravitational field. It is the transmitted projection of the bulk Weyl tensor $C_{ABCD}$ on the brane, such that $E_{AC} = C_{ABCD}n^{B}n^{D}$ with the property, $E_{\mu\nu} = E_{AB}e^{A}_{\mu}e^{B}_{\nu}$. The conservation of matter energy-momentum tensor on the brane i.e $D_{\nu}\tau^{\nu}_{\mu} = 0$, (where $D_{\nu}$ represents the brane covariant derivative) leads to the constraint $D_\nu E^\nu_\mu-\kappa_5^4 D_\nu \pi^\nu_\mu =0$, since $D_\nu \mathcal{F}(R)\delta^\nu_{\mu}=0$ as the bulk Ricci scalar depends only on $\phi$.

The symmetry properties of $E_{\mu\nu}$ allows an irreducible decomposition of the tensor in terms of a given 4-velocity field $u^\mu$ \cite{Maartens:2001jx,Harko:2004ui},
\begin{align}
E_{\mu\nu} &\boldsymbol= -k^{4}\bigg{[}U(r)(u_{\mu}u_{\nu} + \frac{1}{3}\zeta_{\mu\nu})  + 2Q_{(\mu}u_{\nu)}+ P_{\mu\nu}\bigg{]} \label{12}
\end{align}
where $k = \frac{\kappa_{5}}{\kappa_{4}}$ with $\kappa^{2}_{4} = 8\pi G_{4}$ and $\zeta_{\mu\nu} = g_{\mu\nu} + u_{\mu}u_{\nu}$ is the projector orthogonal to $u^\mu$. Note that $\kappa_4^2=\kappa_5^4\lambda_T/6$, such that we retrieve general relativity in the limit $\lambda_T^{-1}\rightarrow 0$ \cite{Harko:2004ui}. In \ref{12} the scalar $U(r) = -\frac{1}{k^{4}}E_{\mu\nu}u^{\mu}u^{\nu}$ is often known as the ``Dark Radiation" term. The second term on the right hand side of \ref{12} consists of a spatial vector $Q_\mu=\frac{1}{k^{4}}\zeta^{\alpha}_{\mu}E_{\alpha\beta}u^{\beta}$ whereas the third term consists of a spatial, tracefree, symmetric tensor $P_{\mu\nu} = -\frac{1}{k^{4}}\big{[}\zeta^{\alpha}_{(\mu}\zeta^{\beta}_{\nu)} - \frac{1}{3}\zeta_{\mu\nu}\zeta^{\alpha\beta}\big{]}E_{\alpha\beta}$.

In order to obtain vacuum solutions on the brane, the brane should be source free such that $\tau_{\mu\nu}=\pi_{\mu\nu}=0$. Thus, the gravitational field equations on the brane reduce to,
\begin{align} 
\mathcal{R}_{\mu\nu} - \frac{1}{2}\mathcal{R}g_{\mu\nu} = -\Lambda_{4}g_{\mu\nu} + \mathcal{F}(R)g_{\mu\nu}- E_{\mu\nu}   \label{EF}
\end{align}
In such a scenario, the effective four-dimensional cosmological constant is given by $\tilde{\Lambda}=\Lambda_4-\mathcal{F(R)}$ while the conservation of energy-momentum tensor on the brane simplifies to, $D_\nu E^\nu_\mu=0$. Additionally, if the solutions are static, the term $Q_\mu$ in \ref{12} should vanish such that the conservation of brane energy-momentum tensor leads to,
\begin{align}
\frac{1}{3}\bar{D}_\mu U +\frac{4}{3}U A_\mu + \bar{D}^\nu P_{\nu\mu} + A^\nu P_{\nu\mu}=0 \label{13}
\end{align}
where $A_\mu=u^\nu D_\nu u_\mu$ is the $4$-acceleration and $\bar{D}$ denotes covariant derivative on the space-like hypersurface orthonormal to $u_\mu$.
Further, if the solutions are spherically symmetric, we may write $A_\mu=A(r)r_\mu$, while the term $P_{\mu\nu}$ can be written as,
\begin{align}
P_{\mu\nu}=P(r)\left(r_\mu r_\nu-\frac{1}{3}\zeta_{\mu\nu}\right) \label{14}
\end{align} 
where $A(r)$ and $P(r)$ (also known as the ``Dark Pressure") are scalar functions of the radial coordinate $r$ and $r_\mu$ is the unit radial vector.

In order to derive static, spherically symmetric solutions of \ref{EF} we consider a metric ansatz of the form,
\begin{align}
 ds^2=-e^{\nu(r)}dt^2 + e^{\lambda(r)}dr^2 + r^2(d\theta^2 +sin^2 \theta d\phi^2 ) \label{15}
\end{align}
and solve for $\nu(r)$, $\lambda(r)$, $U(r)$ and $P(r)$ since \ref{15} satisfies \ref{EF} and \ref{13}. One can show that the solution of these equations lead to the following form for $e^{-\lambda}$ \cite{Chakraborty:2014xla},
\begin{align}
e^{-\lambda} &= 1 - \frac{\Lambda_{4} - \mathcal{F}(R)}{3}r^{2} - \frac{Q(r)}{r} - \frac{C}{r} \label{16}
\end{align}
where $C$ is an arbitrary integration constant and $Q(r)$ is defined as,
\begin{align}
Q(r) = \frac{3}{4\pi G_{4}\lambda_{T}}\int r^{2}U(r)dr \label{16}
\end{align}
From the form of $e^{-\lambda}$ it can be inferred that $Q(r)$ is the gravitational mass originating from the dark radiation and can be interpreted as the ``dark mass" term. It is important to emphasize that in the limit $f(R)\rightarrow R$, $\Lambda_{4} \rightarrow 0$ and $U \rightarrow 0$, we get back the standard Schwarzschild solution and the constant of integration can then be identified with $C = 2G_{4}M$, where $M$ is the mass of the gravitating body.

Further, one can show that for a static, spherically symmetric spacetime the ordinary differential equations for dark radiation $U(r)$ and dark pressure $P(r)$ satisfy \cite{Chakraborty:2014xla},
\begin{align}
\frac{dU}{dr} = -2\frac{dP}{dr} - 6\frac{P}{r} - \frac{(2U + P)[2G_{4}M + Q + \{ \alpha(U + 2P) + 2\chi /3\}r^{3}]}{r^{2}\big{(} 1 - \frac{2G_{4}M}{r} - \frac{Q(r)}{r} - \frac{\Lambda_{4} - \mathcal{F}(R)}{3}r^{2}\big{)}} \label{17}
\end{align}
and
\begin{align}
\frac{dQ}{dr} = 3\alpha r^{2}U \label{18}
\end{align}
where $\alpha = \frac{1}{4\pi G_{4}\lambda_{T}}$ and $\chi = -\tilde{\Lambda}=\mathcal{F}(R) - \Lambda_{4}$. 
\ref{17} and \ref{18} can be recast into a more convenient form namely,
\begin{align}
\frac{d\mu}{d\theta} &= -(2\mu + p)\frac{\tilde{q} + \frac{1}{3}(\mu + 2p) + \frac{l}{3}}{1 - \tilde{q} + \frac{l}{6}} - 2\frac{dp}{d\theta} + 2\mu - 2p \label{19}\\
\frac{d\tilde{q}}{d\theta} &= \mu - \tilde{q} \label{20} 
\end{align} 
by defining the variables, 
\begin{align}
\tilde{q} = \frac{2G_{4}M + Q}{r} ;~~ ~~~ \mu = 3\alpha r^{2}U ; ~~~~~ p= 3\alpha r^{2}P ; ~ ~~~~\theta = ln ~ r ; ~~~ ~~2\chi r^{2} = l
\end{align}
\ref{19} and \ref{20} can be referred to as the differential equations governing the source terms on the brane. 
For a detailed derivation of the differential equations for the metric components and the source terms one is referred to \cite{Chakraborty:2014xla,Harko:2004ui}.
In the next section we shall review various static, spherically symmetric and vacuum solutions of \ref{EF} on the brane.

\section{Various classes of solutions on the brane}\label{S3}
The source equations \ref{19} and \ref{20} for dark radiation and dark pressure cannot be solved simultaneously until we impose some further conditions on them. Hence, we choose some specific relations between dark radiation $U$ and dark pressure $P$, necessarily defining the various equations of state in the framework of the brane world model. We will note that the different choices of equations of state will lead to very distinct solutions.
\subsection{Case A: P = 0}
This is the vanishing dark pressure case. The dark radiation and the dark mass can be evaluated by solving the coupled equations \ref{19} and \ref{20}. With $P=0$, these two equations simplify to,
\begin{align}
\frac{d\tilde{q}}{d\theta} &= \mu - \tilde{q} \rm ~and \label{21}\\
\frac{d\mu}{d\theta} &= 2\mu\bigg{[}\frac{6 - l - 2\mu - 12\tilde{q}}{6 + l - 6\tilde{q}}\bigg{]} \label{22}
\end{align}
respectively. The above two equations can be combined to produce a single differential equation given by,
\begin{align}
(6 + l - 6\tilde{q})\frac{d^{2}\tilde{q}}{d\theta^{2}} + (26\tilde{q} - 6 + 3l)\frac{d\tilde{q}}{d\theta} + 4\bigg{(}\frac{d\tilde{q}}{d\theta}\bigg{)}^{2} + 2\tilde{q}(14\tilde{q} - 6 + l) = 0 \label{23}
\end{align}
Since $l$ is not a constant in \ref{23} we apply some approximate methods to find a solution for $\tilde{q}(\theta)$. By taking Laplace transformation of \ref{23} and using the convolution theorem we get an integral solution for $\tilde{q}(\theta)$, 
\begin{align}
\tilde{q}(\theta) = \tilde{q}_{0}(\theta) + \int^{\theta}_{\theta_{0}} g(\theta - x)\bigg{[}3\tilde{q}\frac{d^{2}\tilde{q}}{dx^{2}} - 13\tilde{q}\frac{d\tilde{q}}{dx} - 2\bigg{(}\frac{d\tilde{q}}{dx}\bigg{)}^{2} - 14\tilde{q}^{2} - \chi e^{2x}\frac{d^{2}\tilde{q}}{dx^{2}} - 3\chi e^{2x}\frac{d\tilde{q}}{dx} - 2\chi e^{2x}\tilde{q}\bigg{]}dx \label{24}
\end{align}
with the associated functions,
\begin{align}
&g(\theta - x) = \frac{1}{9}\big{[}e^{2(\theta - x)} - e^{-(\theta - x)}\big{]} \label{25}\\
&\tilde{q}_{0}(\theta) = B_{1}e^{-\theta} + B_{2}e^{2\theta} \label{26}\\
&B_{1} =[3\tilde{q}(\theta_{0}) - \mu(\theta_{0})]\frac{e^{\theta_{0}}}{3}=M_0-\alpha U(r_0)r_0^3 \label{27}\\
&B_{2} = \mu(\theta_{0})\frac{e^{-2\theta_{0}}}{3}=\alpha U(r_0)  \label{28}
\end{align}
where $\theta_{0} = ln ~ r_{0}$ is an arbitrary point which can be associated with the vacuum boundary of a compact astrophysical object \cite{Chakraborty:2014xla,Harko:2004ui} and $M_0=2G_4 M+Q(r_0)$.

\ref{24} can be solved by applying successive approximation methods. The zeroth order solution denoted by $\tilde{q}_0$ is derived by considering only the linear part of \ref{23}. The full solution can thus be expressed as $\tilde{q}(\theta)=$lim$_{m\rightarrow \infty}\tilde{q}_m (\theta)$, ($m \in N$ being the order of the equation) such that the iterative solution at $m^{th}$ order is connected to the $(m-1)^{th}$ order by the following differential equation \cite{Chakraborty:2014xla,Harko:2004ui},
\begin{align}
\tilde{q}_{m}(\theta) &= \int^{\theta}_{\theta_{0}}H(\theta - x)\bigg{[}3\tilde{q}_{m-1}\frac{d^{2}\tilde{q}_{m-1}}{dx^{2}} - 13\tilde{q}_{m-1}\frac{d\tilde{q}_{m-1}}{dx} - 2\bigg{(}\frac{d\tilde{q}_{m-1}}{dx}\bigg{)}^{2} \nonumber\\
&-14\tilde{q}^{2}_{m-1} - \chi e^{2x}\frac{d^{2}\tilde{q}_{m-1}}{dx^{2}} - 3\chi e^{2x}\frac{d\tilde{q}_{m-1}}{dx} - 2\chi e^{2x}\tilde{q}_{m-1}\bigg{]}dx + \tilde{q}_{m-1}(\theta) \label{29}
\end{align}

Once we determine the solution for $\tilde{q}(\theta)$ we can derive the solution for the metric components by using the gravitational field equations on the brane and the condition for conservation of energy-momentum tensor. In the zeroth order, the static and spherically symmetric solution to the field equations is given by \cite{Chakraborty:2014xla,Harko:2004ui},
\begin{align}
U &= \frac{B_{2}}{\alpha} \label{32} \\
e^{\nu} &= C_{0}\sqrt{\frac{\alpha}{B_{2}}} \label{30}\\
e^{-\lambda} &= 1 - \frac{B_{1}}{r} - B_{2}r^{2} \label{31}
\end{align}
where $C_{0}$ is an arbitrary constant of integration. Since $\alpha$ is positive \ref{30} implies that $B_2$ and consequently $U(r_0)$ should be positive. Also, the $g_{tt}$ component of the metric should be positive, which implies $C_0>0$. 

Iterating once more, we get the approximate expressions for $U(r)$ and $e^{\nu(r)}$ upto first order,
\begin{align}
&U=e^{-2\nu(r)}  \label{33} \\
&e^{\nu(r)}= C_{0}\sqrt{\frac{\alpha}{B_{2}}} + \sqrt{\frac{\alpha r_{0}}{2}}\sqrt{\frac{r}{B_{2}(r_{0} - r)\big{[}B_{1} + B_{2}rr^{2}_{0} + B_{2}r_{0}r^{2}\big{]} + \frac{1}{3}B_{2}\chi rr_{0}(r^{2}_{0} - r^{2})}} \rm  \label{34} 
\end{align}
Since we are interested in the distances much smaller compared to the cosmological horizon $r_0$, it is reasonable to assume $r\ll r_0$. Under this assumption \ref{34} simplifies considerably,
\begin{align}
e^{\nu(r)} \simeq C_{0}\sqrt{\frac{\alpha}{B_{2}}} + \sqrt{\frac{\alpha}{2B_{1}B_{2}}}\rho^{-1/2}\big[1 + \frac{1}{\rho r}\big]^{-1/2} \label{35}
\end{align}
where $\rho = \frac{r_{0}^{2}}{B_{1}}\big(B_{2} + \frac{\chi}{3}\big)$. It is evident from \ref{35} that $B_{2} + \frac{\chi}{3}$ should be positive while $B_1$ can assume both signatures. Further, if $r>1/\rho$ we can perform a binomial expansion of \ref{35} giving rise to a solution of the form,
\begin{align}
e^{\nu(r)} \simeq \delta + \beta - \frac{\beta}{2\rho r} \label{36}
\end{align}
where $\delta = C_{0}\sqrt{\frac{\alpha}{B_{2}}}$ and $\beta = \frac{1}{r_{0}}\sqrt{\frac{\alpha}{2B_{2}^{2} + \frac{2B_{2}\chi}{3}}}$. Note that the dependence on $f(\mathcal{R})$ gravity comes from the parameter $\chi$. \ref{36} can be rescaled such that the $g_{tt}$ component of the metric assumes the form,
\begin{align}
e^{\nu(r)} \simeq 1 - \frac{r_{1}}{2r} \frac{1}{C_{0}r_{0}(2B_{2} + \frac{2\chi}{3})^{1/2} + 1} \simeq 1-\frac{2G_4\tilde{M}}{r}\label{37}
\end{align}
where 
\begin{align}
r_1=\frac{1}{\rho}=\frac{M_{0} - \alpha U(r_{0})r_{0}^{3}}{r_{0}^{2}(\alpha U(r_{0}) + \chi/3)} \label{38}
\end{align}
Therefore it is clear from \ref{37} that in the regime $r_1\ll r \ll r_0$, the $g_{tt}$ component of the approximate metric is very similar to the Schwarzschild spacetime in \gr, although the ADM mass $\tilde{M}$ has contributions from the inertial mass $M$ as well as the higher curvature and higher dimension terms. The $g_{tt}$ component solely determines the photon sphere $r_{ph}$ and the radius of the marginally stable circular orbit $r_{ms}$ of massive test particles. The photon sphere $r_{ph}$ is obtained from the solution of,
\begin{align}
2g_{tt}-rg_{tt,r}=0 \label{40}
\end{align}
while the marginally stable circular orbit $r_{ms}$ is evaluated from the solution of 
\begin{align}
r g_{tt} g_{tt,_{rr}}=2 r g_{tt,_r}^2 -3 g_{tt} g_{tt,_r} \label{41}
\end{align}
Note that $\tilde{M}$ should be positive, otherwise $r_{ph}$ and $r_{ms}$ becomes negative, which is unphysical. Since $\tilde{M}$ and $C_0$ are both positive, together they ensure that $B_1>0$.

In order to simplify our calculations we scale the radial distance $r$ in units of the gravitational radius $r_g=G_4\tilde{M}/c^2$, such that \ref{37} assumes the form,
\begin{align}
e^{\nu(r)} \simeq 1 - \frac{2}{\tilde{r}} \label{37-a}
\end{align}
where $\tilde{r}=r/r_g=r/\tilde{M}$ (with $G_4=c=1$).
The deviation of the approximate metric from the Schwarzschild spacetime is manifested in the $g_{rr}$ term, where
\begin{align}
g^{rr} = e^{-\lambda(r)} \simeq 1 - \frac{\tilde{\varepsilon}}{\tilde{r}} - 3\tilde{\gamma} \tilde{r} + \tilde{\eta} \tilde{r}^{2}+ \tilde{\sigma}\tilde{r}^{4} \label{39}
\end{align}
where,
\begin{align*}
\tilde{\varepsilon}=\frac{\varepsilon}{r_g} &= \frac{1}{r_g}\Bigg\lbrace\bigg[ M_{0} - \frac{\alpha U(r_{0})r_{0}^{3}}{5}\bigg] \bigg[ 1 - \alpha U(r_{0})r_{0}^{2}\bigg] - \frac{4}{5}\alpha U(r_{0})r_{0}^{3}  \bigg[ 1 + \frac{\chi r_{0}^{2}}{3}\bigg] \Bigg\rbrace\\
\tilde{\gamma}=\gamma r_g &= B_{1}B_{2}r_g \\
\tilde{\eta}=\eta r_g^2&= r_g^2\Bigg\lbrace\bigg[ B_{2} + \frac{\chi}{3}\bigg]\bigg[ 1 - 2B_{2}r_{0}^{2}\bigg] - 2B_{2}\bigg[ 1 - \frac{B_{1}}{r_{0}}\bigg]\Bigg\rbrace \\
\tilde{\sigma}=\sigma r_g^4 &= r_g^4\frac{6B_{2}}{5}\big( B_{2} + \frac{\chi}{3}\big)
\end{align*}

Since we are interested in black hole solutions the curvature singularity at $r=0$ must be covered by an event horizon. The radius of the event horizon $r_{EH}$ is obtained from the real positive solutions of $e^{-\lambda(r)}=0$. Since $g^{rr}=0$ is a fifth order algebraic equation, it always has at least one real root. For the real root to be positive we need to choose the values of $\tilde{\varepsilon}$, $\tilde{\gamma}$, $\tilde{\eta}$ and $\tilde{\sigma}$ judiciously. From the previous discussion it is evident that $\tilde{\gamma}$ and $\tilde{\sigma}$ are always positive while 
$\tilde{\varepsilon}$ and $\tilde{\eta}$ can assume any signature. Further constraints on the values of $\tilde{\varepsilon}$, $\tilde{\gamma}$, $\tilde{\eta}$, and $\tilde{\sigma}$ are established from the fact that $r_{EH}<r_{ph}<r_{ms}$.

The disadvantage of this choice of equation of state is that the metric does not represent an exact black hole solution. Following the same iterative procedure we can approximate the metric to second and the next higher orders. However, we have to work out the properties of this metric (namely the $r_{EH}$, $r_{ph}$ and $r_{ms}$) order by order which is not a desirable feature. In the next section we consider another choice of equation of state which will turn out to be more useful.

\subsection{Case B: 2U + P = 0}
In this section we consider an interesting scenario where the dark radiation, ``$U$" and the dark pressure ``$P$" satisfy the constraint $2U+P=0$. For this specific choice, \ref{17} leads to,
\begin{align}
\frac{dP}{dr} = -4\frac{P}{r}. \label{42}
\end{align} 
Therefore the general solution for the dark pressure and the dark radiation is given by,
\begin{align}
P(r) = \frac{P_{0}}{r^{4}}~~~~~{\rm and}~~~~~~
U(r) = -\frac{P_{0}}{2r^{4}} \label{43}
\end{align}
where $P_{0}$ is an arbitrary constant of integration. Consequently, from \ref{18} the dark mass assumes the form,
\begin{align}
Q(r) = Q_{0} + \frac{3\alpha P_{0}}{2r} \label{44}
\end{align}
with the integration constant $Q_{0}$. Using these forms for the source terms the metric components can be computed, where
\begin{align}
e^{\nu(r)} = e^{-\lambda(r)} = 1 - \frac{2G_{4}M + Q_{0}}{r} - \frac{3\alpha P_{0}}{2r^{2}} + \frac{\mathcal{F}(R) - \Lambda_{4}}{3}r^{2} =1-\frac{2G_4\tilde{M}}{r}+\frac{\tilde{\mathcal{Q}}}{r^2}-\frac{\tilde{\Lambda}}{3} r^2\label{45}
\end{align}
This solution is interesting primarily because it represents an exact solution which is very difficult to obtain in the presence of $f(R)$ gravity in higher dimensions. Although \ref{45} resembles the de Sitter/anti-de Sitter Reissner-Nordstr\"{o}m metric in \gr, there are several differences. First, the ADM mass $\tilde{M}$ and the tidal charge parameter $\tilde{\mathcal{Q}}$ have completely different physical origin, i.e. has contributions from the non-local effects of the bulk Weyl tensor which does not happen in \gr. In \ref{45}, $\tilde{\mathcal{Q}}$ can assume both signatures while in \gr\ $\tilde{\mathcal{Q}}$ is always positive. The cosmological constant $\tilde{\Lambda}$ arises naturally in these models and owes its origin to $f(R)$ gravity in higher dimensions. Depending on the relative dominance of $\Lambda_{4}$ and $\mathcal{F}(R)$, $\tilde{\Lambda}$ can be positive, negative or zero, such that the resultant metric is asymptotically de Sitter, anti-de Sitter or flat. 
Recent cosmological observations of distant Type Ia supernovae and the anisotropies in the cosmic microwave background radiation strongly indicate an accelerated expansion of the universe \cite{Bahcall:1999xn,Spergel:2003cb,Spergel:2006hy,Ade:2015rim,Ade:2015xua} which can be explained by a repulsive cosmological constant with positive $\tilde{\Lambda}$. Therefore, it is essential to explore the ramifications of $\tilde{\Lambda}$ in various astrophysical situations. In what follows we will investigate the influence of the cosmological constant in the continuum spectrum emitted by the accretion disk around quasars, which exhibit strong curvature effects near the horizon. Note however, in our case the origin of the cosmological constant is more physically motivated.  

 Again for convenience of future computations we redefine the metric components in terms of the gravitational radius, which for metric \ref{45} is given by $r_g=G_4\tilde{M}$, (with $c=1$) such that the metric components assume the form,
\begin{align}
e^{\nu(r)} = e^{-\lambda(r)} = 1-\frac{2}{\tilde{r}}+\frac{q}{\tilde{r}^2}-\Lambda \tilde{r}^2\label{46}
\end{align}

where $q=\tilde{\mathcal{Q}}/r_g^2$ and $\Lambda=\tilde{\Lambda} r_g^2/3$.

\section{Spectrum from the accretion disk around black holes in the brane embedded in bulk $f(R)$ gravity}\label{S4}
In order to probe the observable effects of $f(R)$ gravity in higher dimensions we consider the near horizon regime of quasars (which host supermassive black holes at the centre) where deviations from \gr\ is expected. The electromagnetic emission from the accretion disk around quasars bears the imprints of the background spacetime and hence can be used as a suitable tool to study the nature of strong gravity.
In this section we compute the signatures of higher dimensional $f(R)$ gravity in the continuum spectrum emitted by the accretion disk around quasars. 

The continuum spectrum of black holes depends not only on the nature of the background spacetime but also on the characteristics of the accretion flow. Depending on the equation of state governing the dark radiation and the dark pressure, the background metric is given by \ref{37} and \ref{39} or \ref{45}.
For the present work we will approximate the accretion flow in terms of the well established ``thin-disk model" \cite{1973blho,Page:1974he} where the accreting fluid is asumed to be confined to the equatorial plane of the black hole such that the resultant accretion disk is geometrically thin with $h(r)\ll r$ ($h(r)$ being the height of the disk at a radial distance $r$).
The azimuthal velocity $u_\phi$ of the accreting fluid dominates the radial velocity $u_r$ and the vertical velocity $u_z$, such that, $u_z \ll u_r \ll u_\phi$. Therefore, such systems do not harbor outflows.  The presence of viscosity reduces the angular momentum of the accreting fluid and generates minimal amount of radial velocity which facilitates slow inspiral and fall of matter into the black hole. The gravitational pull of the black hole is assumed to be much stronger compared to the radial pressure gradients and shear stresses such that the accreting gas falls in nearly circular geodesics.  

The energy-momentum tensor associated with the accreting fluid is given by,
\begin{align}\label{47}
T^{\mu}_ {\nu}= \rho_{0}\left(1+\Pi\right)u^{\mu}u_{\nu}+t^{\mu}_{\nu}+u^{\mu}q_{\nu}+q^{\mu}u_{\nu}
\end{align}
where, $\rho _{0}u^{\mu}u^{\nu}$ is the stress tensor associated with the geodesic flow ($\rho_{0}$ being the proper density and $u^{\alpha}$, the 4-velocity of the accreting fluid), $\Pi\rho _{0}u^{\mu}u^{\nu}$ constitutes the stress-energy tensor from the specific internal energy ($\Pi$) of the system, $t^{\mu\nu}$ represents the energy-momentum tensor evaluated in the local inertial frame of the accreting fluid and $q^{\mu}$ is the heat flux relative to the local rest frame. Note that both $t^{\mu\nu}$ and $q^\mu$ are orthogonal to the 4-velocity, such that ${u^{\nu} t^{\mu}{_\nu}=0=u^\mu q_\mu}$. In the thin-disk approximation, $\Pi\ll 1$ such that the special relativistic correcions to the local hydrodynamic, thermodynamic and radiative properties of the fluid can be safely neglected. Therefore, the entire heat generated due to viscous dissipation is completely radiated away and the accreting fluid retains no heat. As a consequence, only the z-component of the energy flux vector $q^\alpha$ has a non-zero contribution to the stress-energy tensor. For a more elaborate description of the thin accretion disk model one is referred to\cite{1973blho,Page:1974he,Banerjee:2017npv}.

The black hole is assumed to accrete at a steady rate $\dot{M_0}$ and the accreting fluid is assumed to obey conservation of mass, angular momentum and energy. The conservation of mass is given by,
\begin{align}
&\dot{M_0}=-2\pi \sqrt{-g} \Sigma u^r \label{48}
\end{align}
where $g$ represents the determinant of the metric whose effect on the spectrum we intend to study and $\Sigma$ is the surface density of the accreting fluid. The conservation of angular momentum and energy assumes the forms,
\begin{align}\label{49}
\left[\dot{M_0}L-2\pi \sqrt{-g}W^r_\phi \right]_{,r}=4\pi \sqrt{-g} F L
\end{align}
and 
\begin{align}\label{50}
\left[\dot{M_0}E-2\pi \sqrt{-g}\Omega W^r_\phi \right]_{,r}=4\pi \sqrt{-g} F E 
\end{align}
respectively, where $\Omega$ is the angular velocity, $L=u_\phi$ is the specific angular momentum and $E=-u_t$ is the specific energy of the accreting fluid. The flux from the disk is given by $F$ where,
\begin{align}\label{51}
F \equiv \left\langle q^z(r,h)\right\rangle=\left\langle -q^z(r,-h)\right\rangle 
\end{align}
while the height averaged stress tensor in averaged rest frame is denoted by,
\begin{align}\label{52}
\int^h_{-h}dz\left\langle t^\alpha_\beta\right\rangle=W^\alpha_\beta
\end{align}
The conservation laws can be manipulated such that the flux $F(r)$ from the accretion disk is given by,
\begin{align}\label{53}
F=\frac{\dot{M}_0}{4\pi\sqrt{-g}}\tilde{f}
\end{align}
where,
\begin{align}\label{54}
\tilde{f}=-\frac{\Omega_{,r}}{(E-\Omega L)^2}\left[EL-E_{ms}L_{ms}-2\int_{r_{ms}}^r LE_{,r^\prime}dr^\prime \right]
\end{align}
\ref{53} is derived by assuming that the viscous stress $W^r_\phi$ vanishes at the last stable circular orbit such that the accretion disk truncates at $r_{ms}$. After crossing the marginally stable circular orbit the accreting matter falls radially into the black hole. 

By studying geodesic motion of massive test particles in a given static, spherically symmetric spacetime one can derive the angular velocity $\Omega$, the specific energy $E$ and the specific angular momentum $L$ in terms of the metric components, such that 
\begin{align}\label{55}
\Omega=\frac{d\phi}{dt}=\frac{ \sqrt{-\left\lbrace g_{\phi\phi,r}\right\rbrace \left\lbrace g_{tt,r}\right\rbrace}}{g_{\phi\phi,r}}
\end{align}
\begin{align}\label{56}
E=-u_t=\frac{-g_{tt}}{\sqrt{-g_{tt}-\Omega^2 g_{\phi\phi}}} 
\end{align}
and
\begin{align}\label{57}
L=u_\phi=\frac{\Omega g_{\phi\phi}}{\sqrt{-g_{tt}-\Omega^2 g_{\phi\phi}}}
\end{align}

In \ref{54}, $r_{ms}$ represents the radius of the marginally stable circular orbit while $E_{ms}$ and $L_{ms}$ are specific energy and specific angular momentum at $r_{ms}$.
The marginally stable circular orbit is obtained from the point of inflection of the effective potential $V_{eff}$ in which the massive test particles move. Therefore it is obtained from the relation,
$V_{eff}=V_{eff_{,r}}=V_{eff_{,rr}}=0$ where $V_{eff}$ is given by,
\begin{align}\label{58}
V_{eff}(r)=\frac{E^2g_{\phi\phi}+L^2g_{tt}}{-g_{tt}g_{\phi\phi}}-1
\end{align}
Using \ref{56} and \ref{57}, \ref{58} can be simplified to give \ref{41} which can be solved to obtain $r_{ms}$.

The photons thus generated in the system undergo repeated collisions with the accreting gas such that a thermal equilibrium is established between matter and radiation. Such an accretion disk is therefore geometrically thin but optically thick. Consequently, the disk radiates a Planck spectrum at every radial distance $r$ with peak temperature given by $T(r)=\left(\tilde{F}(r)/\sigma\right)^{1/4}$ where $\tilde{F}(r)=F(r)c^6/G_4^2$ (bringing back the $G_4$ and $c$) and $\sigma$ denotes the Stefan Boltzmann constant.
By integrating the Planck function $B_\nu(T(r))$ over the disk surface one can compute the luminosity $L_\nu$ from the disk at an observed frequency $\nu$, such that,
\begin{align}
L_\nu&=8\pi^2 r_g^2\cos i  \int_{r_{\rm ms}}^{r_{\rm out}}\sqrt{-g_{rr}} B_{\nu}(T(\tilde{r}))\tilde{r} d\tilde{r} ~~~{\rm and}\nonumber \\
B_\nu (T)&=\frac{2h\nu^3/c^2}{{\rm exp}\left(\frac{h\nu}{z_g kT}\right)-1} \label{59}
\end{align}
where, $r_g$ denotes the gravitational radius, $i$ represents the inclination angle of the disk to the line of sight and $z_g$ is the 
gravitational redshift factor which relates the modification induced in the photon frequency while travelling from the emitting material to the observer \cite{Ayzenberg:2017ufk}. The gravitational redshift factor is given by,
\begin{align}
z_g=E\frac{\sqrt{-g_{tt}-\Omega^2 g_{\phi\phi}}}{E-\Omega L} \label{60}
\end{align} 
Since the spectrum from the accretion disk is an envelope of a series of black body spectra emitted at different peak temperatures, it is often called a multi-color/multi-temperature black body spectrum.  Note that the theoretical spectrum depends chiefly on the $g_{tt}$ component of the metric while the $g_{rr}$ component is required only during the integration of the flux to obtain the luminosity (see \ref{59}) \cite{Banerjee:2017npv}.

\subsection{Effect of bulk $f(R)$ gravity on the emission from the accretion disk}\label{S4.1}
In the present work we are interested in investigating the modifications induced in the continuum spectrum of quasars due to the presence of $f(R)$ gravity in higher dimensions. The background spacetime is therefore given by \ref{37-a} and \ref{39} for equation of state $P=0$, while \ref{46} denotes the background metric when the equation of state is given by $2U+P=0$.

\begin{figure}[t!]
\centering
\hbox{\hspace{-4.8em}
\includegraphics[scale=0.72]{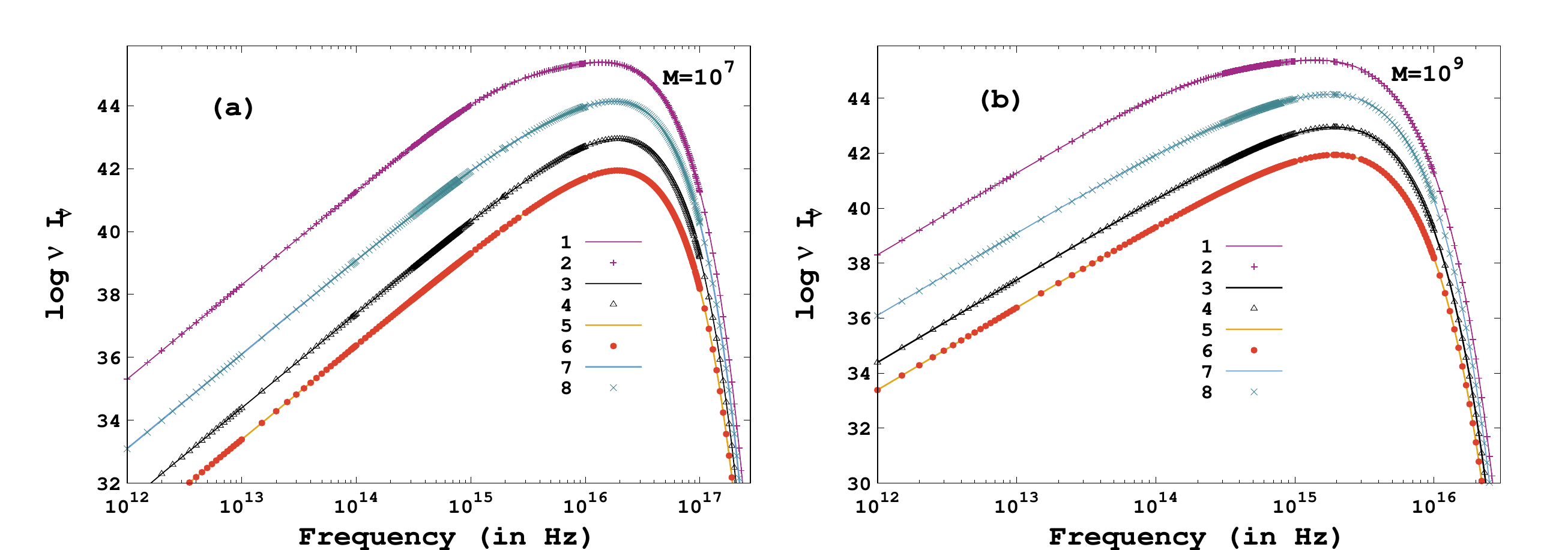}}
\caption{Figure 1: The above figure illustrates variation of the theoretically derived  luminosity from the accretion disk with frequency for two different masses of supermassive black holes. The background is given by \ref{37-a} and \ref{39}.
Both figures 1(a) and 1(b) exhibit a set of eight spectra which are drawn to explain the impact of various metric parameters on the theoretical spectrum. The metric parameters corresponding to spectra ``1"-``8" are reported in \ref{Table1}. 
The accretion rate assumed is $1 M_{\odot}\textrm{yr}^{-1}$ and $\cos i$ is taken to be $0.8$.}
\label{Fig_01}
\end{figure}

In \ref{Fig_01} we plot the theoretically derived spectrum from the accretion disk when the equation of state is given by $P=0$ for two different masses of supermassive black holes, namely, $10^7 M_\odot$ ({\color{blue}Fig. 1(a)}) and $10^9 M_\odot$ ({\color{blue}Fig. 1(b)}). For each of the masses eight spectra $\rm 1-8$ are plotted in \ref{Fig_01} by varying the various metric parameters in \ref{39} which are detailed in \ref{Table1}. In each of the spectra the $g_{tt}$ component is similar to the Schwarzschild spacetime (see \ref{37-a}) while the 
$g_{rr}$ component has several corrections to the Schwarzschild metric (see \ref{39}). 
From \ref{Table1} it is clear that the spectrum labelled by ``1" corresponds to the Schwarzschild scenario although the ADM mass owes its origin to $f(R)$ gravity in higher dimensions in the action. This difference in the origin of mass of the black hole cannot be perceived by an external observer. In spectrum ``2" the space-time is still Schwarzschild-like although the mass term in the $g_{tt}$ and $g_{rr}$ components of the metric are not the same. From \ref{Fig_01} it is clear that this change hardly affects the theoretical spectrum. In spectra ``3" and ``4" the mass term in $g_{rr}$ is same as that of the $g_{tt}$ component while $\tilde{\sigma}$ and $\tilde{\gamma}$ are simultaneously changed as per \ref{Table1}. \ref{Fig_01} shows that change of $\tilde{\sigma}$ has an important effect in the spectrum (since spectra ``1" and ``3" show deviations) while changing $\tilde{\gamma}$ barely has any impact (since spectra ``3" and ``4" are overlapping). For spectra ``5" to ``8" we fix $\tilde{\varepsilon}$ and $\tilde{\gamma}$ since we have understood their effect on the spectrum. Overlap of spectra ``5" and ``6" imply that  $\tilde{\eta}$ has negligible effect on the spectrum. The variation in spectra ``1", ``3" and ``5" are chiefly due to the disparity in the values of $\tilde{\sigma}$. However, once $\tilde{\sigma}$ is lowered below $10^{-6}$ the spectrum becomes insensitive to the changes. This is inferred from the overlap of spectra ``7" and ``8". 

\begin{table}[H]
\vskip0.5cm
{\centerline{\large Table 1}}
{\centerline{Choice of metric parameters corresponding to spectra 1-8 in \ref{Fig_01}}}
\caption{}
\label{Table1}
{\centerline{}}
\begin{center}
\begin{tabular}{|c|c|c|c|c|}

\hline
$\rm Spectrum$ & $\tilde{\varepsilon}$ & $\tilde{\gamma}$ & $\tilde{\eta}$ & $\tilde{\sigma}$ \\
\hline 
$\rm 1$ & $\rm 2.0$ & $\rm 0.0$ & $\rm 0.0$ & $\rm 0.0$ \\ \hline
$\rm 2$ & $\rm 0.5$ & $\rm 0.0$ & $\rm 0.0$ & $\rm 0.0$ \\ \hline
$\rm 3$ & $\rm 2.0$ & $\rm 9.0$ & $\rm 0.0$ & $\rm 1.0$ \\ \hline
$\rm 4$ & $\rm 2.0$ & $\rm ~~10^{-7}$ & $\rm 0.0$ & $\rm 1.0$ \\ \hline
$\rm 5$ & $\rm 2.0$ & $\rm 1.0$ & $\rm ~~100.0$ & $\rm 100.0$ \\ \hline
$\rm 6$ & $\rm 2.0$ & $\rm 1.0$ & $\rm -100.0$ & $\rm 100.0$ \\ \hline
$\rm 7$ & $\rm 2.0$ & $\rm 1.0$ & $\rm 1.0$ & $\rm 10^{-6}$ \\ \hline
$\rm 8$ & $\rm 2.0$ & $\rm 1.0$ & $\rm 1.0$ & $\rm 10^{-10}$ \\ \hline
\end{tabular}

\end{center}
\end{table}

\begin{figure}[t!]
\centering
\hbox{\hspace{6.3em}
\includegraphics[scale=0.72]{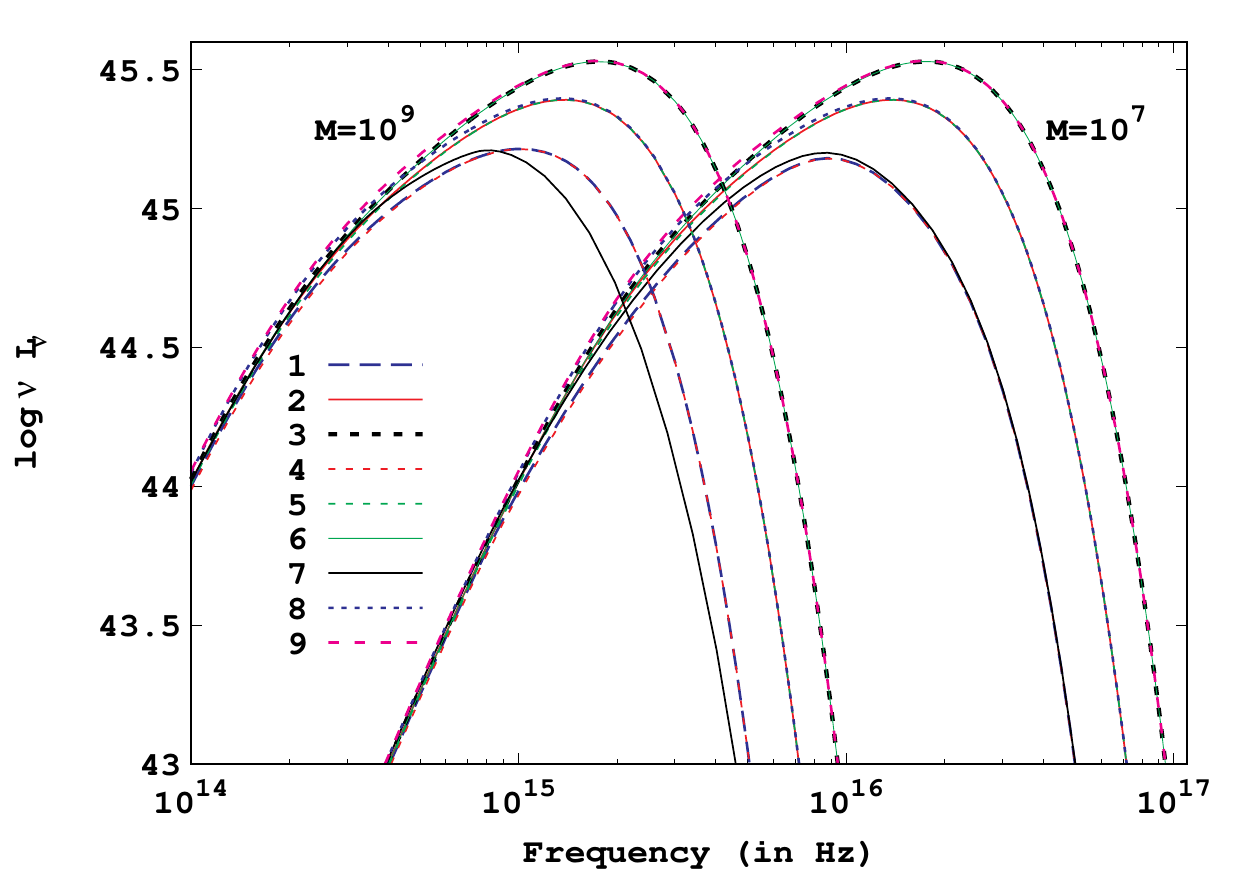}}
\caption{Figure 2: The above figure illustrates the effect of the metric \ref{46} on the 
theoretically derived spectrum from the accretion disk for two different masses of supermassive black holes.  
The accretion rate assumed is $1 M_{\odot}\textrm{yr}^{-1}$ and $\cos i$ is taken to be $0.8$.}
\label{Fig_02}
\end{figure}

\ref{Fig_02} depicts the variation of the theoretically derived luminosity with frequency for black hole masses $10^7 M_\odot$ and $10^9 M_\odot$ when the background spacetime is given by \ref{46} which corresponds to the equation of state $2U+P=0$. 
The values of the metric parameters corresponding to the nine spectra illustrated in \ref{Fig_02} are given in \ref{Table2}. Spectra ``1", ``4" and ``7" corresponds to a constant magnitude of $q=-3$, spectra ``2", ``5" and ``8" are commensurate with $q=0$ while spectra ``3", ``6" and ``9" are in tandem with $q=0.95$. For each set of constant $q$ spectra the cosmological constant $\Lambda$ is variable according to \ref{Table2}.
From the virtual overlap of the spectra with constant $q$ but variable $\Lambda$, it is quite explicit that the tidal charge parameter $q$ has a more significant impact on the spectrum than the cosmological constant $\Lambda$. Only for $q=-3$ the spectrum with $\Lambda < 0$ appears to be deviated from its $\Lambda\geq 0$ counterparts. Note that we cannot choose the magnitude of $\Lambda$ arbitrarily large as this is in odds with the cosmological observations \cite{Spergel:2003cb,Ade:2015xua}. On the other hand if $\Lambda$ is extremely small, it will hardly affect the spectrum. The magnitude of $\Lambda$ should therefore be chosen in an optimal range. 

Moreover, from a theoretical point of view there are restrictions on the maximum positive value of $\Lambda$. 
This stems from the fact that, unlike anti de-Sitter spacetime, a de-Sitter spacetime has a cosmological horizon $r_{CH}$ which is obtained from the largest solution of $e^{-\lambda(r)}=0$ in \ref{46}. Our region of interest $r$ should therefore be confined in the region $r_{EH}<r<r_{CH}$, i.e., the outer radius of the accretion disk $r_{out}$ should be within $r_{CH}$. The fact that the inner radius of the disk $r_{in}$ truncates at $r_{ms}$ automatically ensures that $r_{in}>r_{EH}$. With an enhancement in $\Lambda$, $r_{CH}$ shrinks while $r_{EH}$ increases, such that for $\Lambda=\Lambda_{max}=1/27$ (and $q=0$) the two horizons coincide and for higher values of $\Lambda$, the horizons disappear leading to the formation of a naked singularity \cite{Stuchlik:2008ea}. The presence of $q$ slightly modifies $\Lambda_{max}$ with a negative $q$ marginally lowering the value as opposed to a positive $q$. Also note that we cannot arbitrarily increase $q$, once again to preserve the cosmic censorship conjecture. In the absence of $\Lambda$, the presence of an event horizon requires $q\leq 1$. On increasing the negative value of $\Lambda$, the maximum value of $q$ gets marginally lowered (e.g. $q_{max}\sim 0.925$ if $\Lambda \sim -0.1$) while the presence of a de Sitter $\Lambda$ enhances the $q_{max}$ (e.g. $q_{max}\sim 1.01$ if $\Lambda \sim 0.05$). However, no real value of $\Lambda$ can raise $q_{max}$ upto $1.1$. Therefore, for all practical purposes we will confine ourselves to $\Lambda <1/27$ and $q\leq 1$. 

\begin{table}[H]
\vskip0.5cm
{\centerline{\large Table 2}}
{\centerline{Choice of metric parameters corresponding to spectra 1-9 in \ref{Fig_02}}}
\caption{}
\label{Table2}
{\centerline{}}
\begin{center}
\begin{tabular}{|c|c|c|}
\hline
$\rm Spectrum$ & $q$ & $\Lambda$  \\ 
\hline
$\rm 1$ & $\rm -3.0$ & $\rm 0.0$ \\ \hline
$\rm 2$ & $\rm  ~~0.0$ & $\rm 0.0$ \\ \hline
$\rm 3$ & $\rm  ~~~0.95$ & $\rm 0.0$ \\ \hline
$\rm 4$ & $\rm -3.0$ & $\rm ~~~~~~7\times 10^{-9}$ \\ \hline
$\rm 5$ & $\rm  ~~0.0$ & $\rm ~~~~~~7\times 10^{-9}$ \\ \hline
$\rm 6$ & $\rm ~~~0.95$ & $\rm ~~~~~~7\times 10^{-9}$ \\ \hline
$\rm 7$ & $\rm -3.0$ & $\rm ~~-2\times 10^{-7}$ \\ \hline
$\rm 8$ & $\rm ~~0.0$ & $\rm ~~-2\times 10^{-7}$ \\ \hline
$\rm 9$ & $\rm ~~~0.95$ & $\rm ~~-2\times 10^{-7}$ \\ \hline
\end{tabular}

\end{center}
\end{table}

A more stringent constraint on $\Lambda_{max}$ is established from the fact that no stable circular orbit exists for $\Lambda>2.37\times 10^{-4}$ in the absence of the charge parameter \cite{Stuchlik:2008ea}. Once again the presence of a negative $q$ further lowers $\Lambda_{max}$ while a positive $q$ raises this value upto a maximum of $\Lambda\sim 7\times 10^{-4}$. Since our accretion disk truncates at $r_{ms}$ we need to keep the maximum value of $\Lambda$ well below $2.37\times 10^{-4}$.\\ 
The choice of $\Lambda$ automatically restricts the maximum extent of the accretion disk.
This is because a positive $\Lambda$ has a repulsive effect as opposed to the attractive force offered by the central black hole. Therefore, the physically relevant region for accretion is the regime where the attractive force due to the black hole dominates. This is given by the static radius $r_s$ where the attractive force due to the black hole and the repulsive force due to $\Lambda$ nullify. The value of $r_s$ diminishes with an increase in $\Lambda$ and is evaluated from the turning point of the pseudo-Newtonian potential $\Psi$ experienced by the test particles while moving in a given spacetime, in our case \ref{46}, where, 
\begin{align}
\Psi=\int dr \frac{L^2}{E^2}r^3 \label{61} 
\end{align}    
and $E$ and $L$ are given by \ref{56} and \ref{57}. The outer radius of the accretion disk $r_{out}$ must be lesser than $r_s$ for accretion to take place. 

In this work we will take $r_{out}\sim 500 R_g$ which is just a typical choice \cite{Walton:2012aw,Bambi:2011jq}. This further brings down the maximum allowed value for $\Lambda\leq 7\times 10^{-9}=\Lambda_{max}$.  
Although the outer radius of the accretion disk can deviate from our choice it will not affect the results substantially since a greater $r_{s}$ (and hence a larger $r_{out}$) will diminish $\Lambda_{max}$ by orders of magnitude which will have negligible effect on the spectrum. A smaller $r_{out}$ on the other hand will increase $\Lambda_{max}$ but the effective disk luminosity will not change much since the flux is integrated over a smaller area of the disk. In fact one can verify that by raising $\Lambda_{max}\sim 10^{-6}$ the deviation from the the Schwarzschild/\RN\ scenario \cite{Stuchlik:2008ea,Perez:2012bx} is minimal. 
Since the magnitude of $\Lambda$ is very small one needs to choose $r_{out}$ judiciously 
in order to detect an observable effect of the cosmological constant on the spectrum.

A feature common to both \ref{Fig_01} and \ref{Fig_02} is that the disk luminosity of a lower mass black hole peaks at a higher frequency. This is because the peak temperature of the local black body emission is inversely proportional to the mass, $T(r)\propto M^{-1/4}$ (see discussion above \ref{59}). Hence disk emission from stellar mass black holes peak in soft X-rays while for supermassive black holes the maximum emission occurs in the optical domain.  
We also note that the spectra in \ref{Fig_01} and \ref{Fig_02} are different in the sense that the deviation from GR shows up in the low energy domain in \ref{Fig_01} and high energy regime in \ref{Fig_02}. This is attributed to the fact that the $g_{rr}$ term of the background metric governing \ref{Fig_01} has a $\tilde{\sigma}\tilde{r}^4$ contribution in the denominator which suppresses the luminosity from the Schwarzschild scenario. It is evident from \ref{Fig_01} that even a minimal deviation of $\tilde{\sigma}\sim 10^{-6}$ causes a substantial departure from the general relativistic counterpart (see \ref{Table1} and \ref{Fig_01}). The $\tilde{r}^4$ dependence of the $g_{rr}$ component of the metric ensures that the outer disk which emits in lower frequencies has the dominant contribution in the luminosity. Hence, the deviation from \gr\ in \ref{Fig_01} becomes evident in the lower frequencies. On the contrary, the metric components corresponding to \ref{Fig_02} have inverse powers of $\tilde{r}$, hence deviations from GR are manifested chiefly in the inner disk which emits high energy radiations.

\section{NUMERICAL ANALYSIS }\label{S5}
In this section we use the thin-disk approximation for the accretion flow in the background spacetime given by \ref{46}, since this represents an exact black hole solution, to evaluate the theoretical estimates of optical luminosities for a sample of eighty Palomar Green (PG) quasars \cite{Schmidt:1983hr,Davis:2010uq}. We compute $L_{opt}\equiv \nu L_\nu$ at the wavelength 4861\AA\ following Davis \& Laor \cite{Davis:2010uq}. 
The masses of these quasars have been determined previously by the method of reverberation mapping \cite{Kaspi:1999pz,Kaspi:2005wx,Boroson:1992cf,Peterson:2004nu} and for a small sub-sample of thirteen quasars the masses are also known by the $M-\sigma$ method \cite{Ferrarese:2000se,Gebhardt:2000fk,Tremaine:2002js}.
The bolometric luminosities of these quasars have been estimated using observed data in the optical \cite{1987ApJS...63..615N}, UV \cite{Baskin:2004wn}, far-UV \cite{Scott:2004sv}, and soft X-ray \cite{Brandt:1999cm} domain. For all the quasars in the sample, the accretion rates and the observed estimates of the optical luminosity are reported in \cite{Davis:2010uq}. Since we are modelling the accretion disk of quasars whose emission peaks in the optical part of the spectral energy distribution (SED), we are primarily interested in accurate and precise estimates of the optical luminosity. 

In order to compute the theoretical optical luminosity the inclination of the accretion disk ``$i$" is also required (\ref{59}). For quasars ``$cosi$" generally ranges from $0.5-1$ since emissions from nearly edge-on systems are likely to be obscured. This permits us to neglect the effect of light bending while computing the spectrum from the accretion disk. Such effects become  conspicuous for disks with high inclination angles \cite{Bambi:2012tg,Kong:2014wha}.
In this work, we assume a typical value of $cosi \sim 0.8$ for all the quasars \cite{Davis:2010uq}. This is further supported from the fact that the error (e.g., reduced $\chi^2$, Nash-Sutcliffe efficiency, index of agreement etc.) between the theoretical and observed luminosities for non-rotating black holes with a fixed $q$ gets minimized when cosi lies between $0.77- 0.82$ \cite{Banerjee:2017hzw}. The inclination angles of some of the quasars in our sample have been independently determined by Piotrovich et al. \cite{2017Ap&SS.362..231P} by estimating the degree of polarisation of the scattered radiation from the accretion disk. It turns out that their estimates are consistent with our choice.

In order to understand whether the presence of bulk $f(R)$ gravity provides a better approximation to the observed spectra we calculate the theoretical estimates of the optical luminosity $L_{opt}$ for all the eighty quasars with known masses, accretion rates and disk inclination. This is compared with the corresponding observed values $L_{obs}$ to deduce the most favored choice of the metric parameters ($q$ and $\Lambda$) that explains observations the best. To arrive at the preferred model for $q$ and $\Lambda$ we discuss several error estimators:

\begin{itemize}
\item {\bf Chi-square $\boldsymbol {\chi^{2}}~$}:~
If $\{ \mathcal{O}_{i}\}$ represents a set of observed data with possible errors $\{ \sigma_{i} \}$, and $\Omega_{i}(q, \Lambda)$ denotes the corresponding model estimates of the observed quantity with model parameters $q$ and $\Lambda$, then the $\chi^{2}$ of the distribution is given by,
\begin{align}
\chi^{2} (q, \Lambda) = \sum_{i} \frac{\{ \mathcal{O}_{i} - \Omega_{i}(q, \Lambda) \}^{2}}{\sigma_{i}^{2}} \label{61}
\end{align}
For our sample, the error $\{ \sigma_{i} \}$ corresponding to optical luminosities of individual quasars are not reported. Hence we assign equal weightage to every observation.
\begin{figure}[t!]
\centering
\includegraphics[scale=0.75]{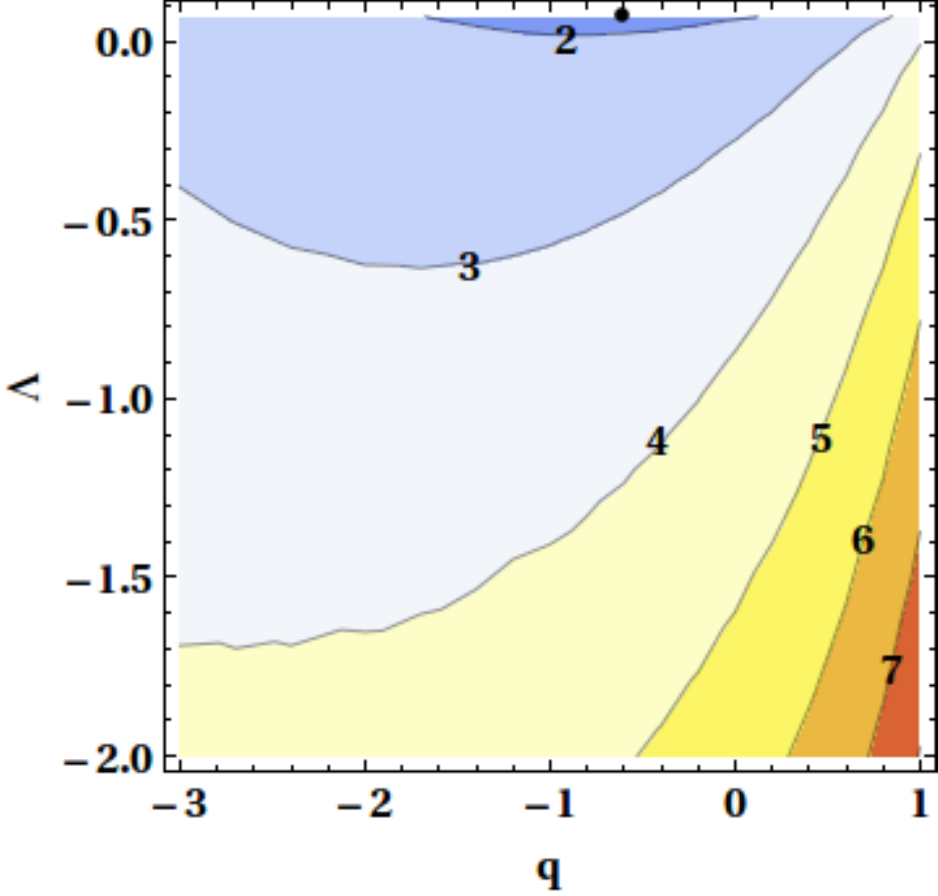}
\caption{Figure 3: The above figure depicts contours of constant $\chi^{2}$ as a function of the metric parameters $q$ and $\Lambda$. The values of $\Lambda$ are in the units of $10^{-7} r_g^{-2}$. The minimum of the $\chi^{2}$ is denoted by the black dot. It is evident from the plot that $\chi^{2}$ minimizes for a negative value of $q \sim -0.6$ and a positive $\Lambda \sim \mathcal{O}(10^{-9}) $.   
While a negative $q$ marks a clear deviation from \gr, a positive $\Lambda$ indicates that a de-Sitter spacetime is favored by electromagnetic observations from quasars. }
\label{Fig_3}
\end{figure}
The values of $q$ and $\Lambda$ that minimize $\chi^{2}$ represent the most favored values of the metric parameters. 

It is interesting to note that although $\chi^{2}$ turns out to be a valid error estimator, reduced chi-square $\chi ^{2}_{Red}=\chi ^{2}/\nu$, (with $\nu$ being the degrees of freedom) is not useful in our case since the number of degrees of freedom for our model is not very well-defined. This is attributed to the fact that there are restrictions to the values of both $q$ and $\Lambda$ (see \ref{S4.1}).  Such systems are known as models with prior where definition of degrees of freedom requires additional inputs apart from the number of parameters in the model \cite{Andrae:2010gh}. 

\ref{Fig_3} shows the constant $\chi^{2}$ contours for different values of the metric parameters $q$ and $\Lambda$. The values of the brane cosmological constant $\Lambda$ are expressed in units of $10^{-7}  r_g^{-2}$. From the figure it is clear that $\chi^{2}$ achieves a minimum value $\sim 1.78$ (denoted by the black dot) for a negative tidal charge parameter $q \sim -0.6$ and a positive $\Lambda\sim 7\times 10^{-9}$. Since \gr$~$ cannot account for a negative tidal charge parameter, this may signal higher dimensions at play in the strong gravity regime around quasars. Note that the signature of $q$ is more important than its exact value since negative tidal charge parameters does not arise in \gr. A positive $\Lambda$ on the other hand signifies that a de-Sitter spacetime is preferred from the continuum spectra of quasars. This is in agreement with the cosmological observations \cite{Spergel:2003cb,Ade:2015xua}. In the next section we will comment on how the value of $\Lambda$ estimated from our analysis compares with the cosmological constant measured from observations related to distant Type Ia supernovae and cosmic microwave background radiation. Before that we discuss a few more error estimators to confirm the robustness of our results. 

\pagebreak
\item \textbf{Nash-Sutcliffe Efficiency and its modified form:}
\begin{figure}[htp]
\subfloat[Nash Sutcliffe Efficiency \label{Fig_4a}]{\includegraphics[scale=0.65]{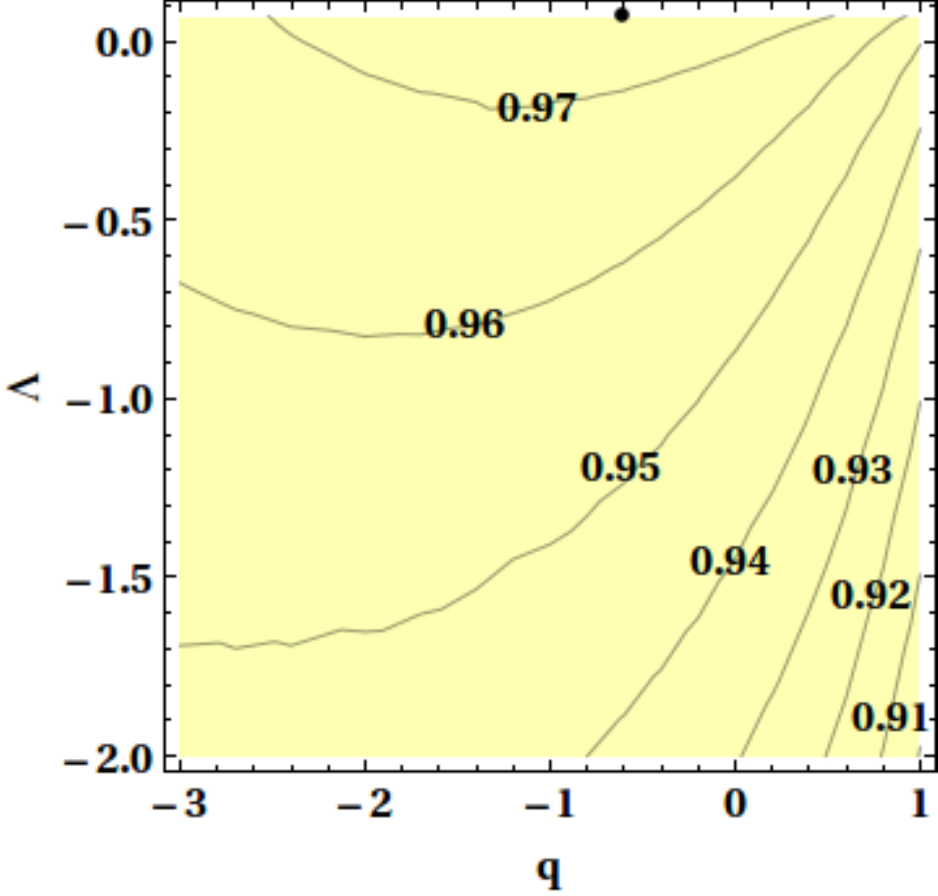}}
\hspace{3cm}
\subfloat[Modified Nash Sutcliffe Efficiency\label{Fig_4b}]{\includegraphics[scale=0.65]{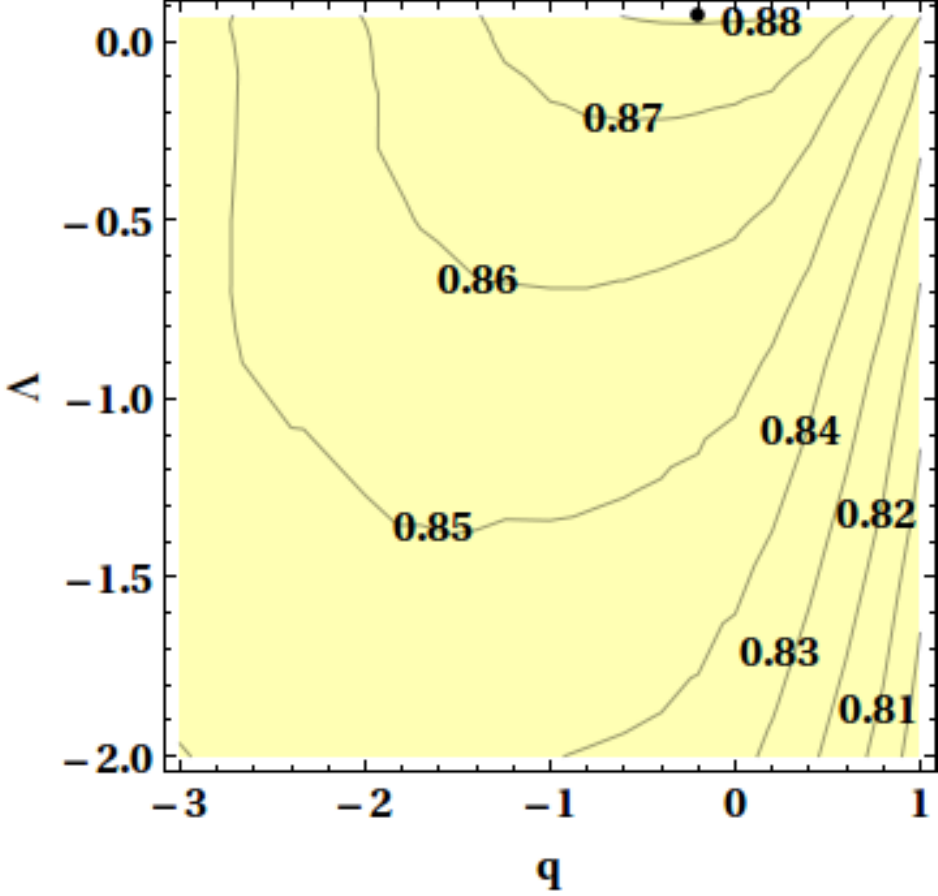}}
\caption{Figure 4: The above figure illustrates contours of constant (a) Nash-Sutcliffe Efficiency $E$ and (b) the modified form of the Nash-Sutcliffe Efficiency $E_1$ with the tidal charge parameter $q$ and $\Lambda$. As before, the values of $\Lambda$ are in the units of $10^{-7} r_g^{-2}$. Both the error estimators maximize for negative values of $q$ and positive $\Lambda\sim \mathcal{O}(10^{-9})$.}
\end{figure}
Nash-Sutcliffe Efficiency $E$ \cite{NASH1970282,WRCR:WRCR8013,2005AdG.....5...89K} is given by,
\begin{align}
E(q, \Lambda) = 1 - \frac{\sum_{i} \{ \mathcal{O}_{i} - \Omega_{i}(q, \Lambda) \}^{2}}{\sum_{i} \{ \mathcal{O}_{i} - \mathcal{O}_{av} \}^2} \label{62}
\end{align}
It relates the sum of the absolute squared differences between the theoretical predictions $\Omega_{i}$ and the observed values $\mathcal{O}_{i}$, normalized by the variance of the observed values. In \ref{62} $\mathcal{O}_{av}$ denotes the mean observed optical luminosity of the quasars. $E$ can assume a maximum value of 1. A model with $E \sim 1$ is ideal since it predicts the observations with greatest accuracy. From \ref{62} it is clear that $E$ can acquire negative values and may go upto $-\infty$. A model with negative $E$ indicates that the average of the observed data is a better predictor than the model.
\par
Since Nash-Sutcliffe Efficiency $E$ is susceptible to be oversensitive to higher values of the luminosity, a modified version of the same is proposed which is denoted by $E_1$ \cite{WRCR:WRCR8013}. This is due to the presence of the square of the error in the numerator in \ref{62}. Accordingly, the modified Nash-Sutcliffe Efficiency $E_1$ is defined to be,
\begin{align}
E_{1}(q, \Lambda) = 1 - \frac{\sum_{i}|\mathcal{O}_{i} - \Omega_{i}(q, \Lambda)|}{\sum_{i}|\mathcal{O}_{i} - \mathcal{O}_{av}|} \label{63}
\end{align}
such that it succeeds to enhance the sensitivity of the estimator towards lower values of optical luminosity. Similar to $E$, the most favored model of $q$ and $\Lambda$ should maximize $E_1$. 

In \ref{Fig_4a} and \ref{Fig_4b} we plot contours of constant $E$ and $E_1$ respectively, as functions of $q$ and $\Lambda$. As before, the black dot in both the figures indicate the coordinates of maximum $E$ and $E_1$. The figures explicitly elucidate that 
both the error estimators maximize for negative $q$ and positive $\Lambda\sim\mathcal{O}(10^{-9})$ which is in agreement with our previous findings. This may be an indication of some new physics at play in the strong gravity regime, higher dimensions being one such possibility.

\item \textbf{Index of agreement and its modified form:} 
\begin{figure}[htp]
\subfloat[Index of agreement \label{Fig_5a}]{\includegraphics[scale=0.65]{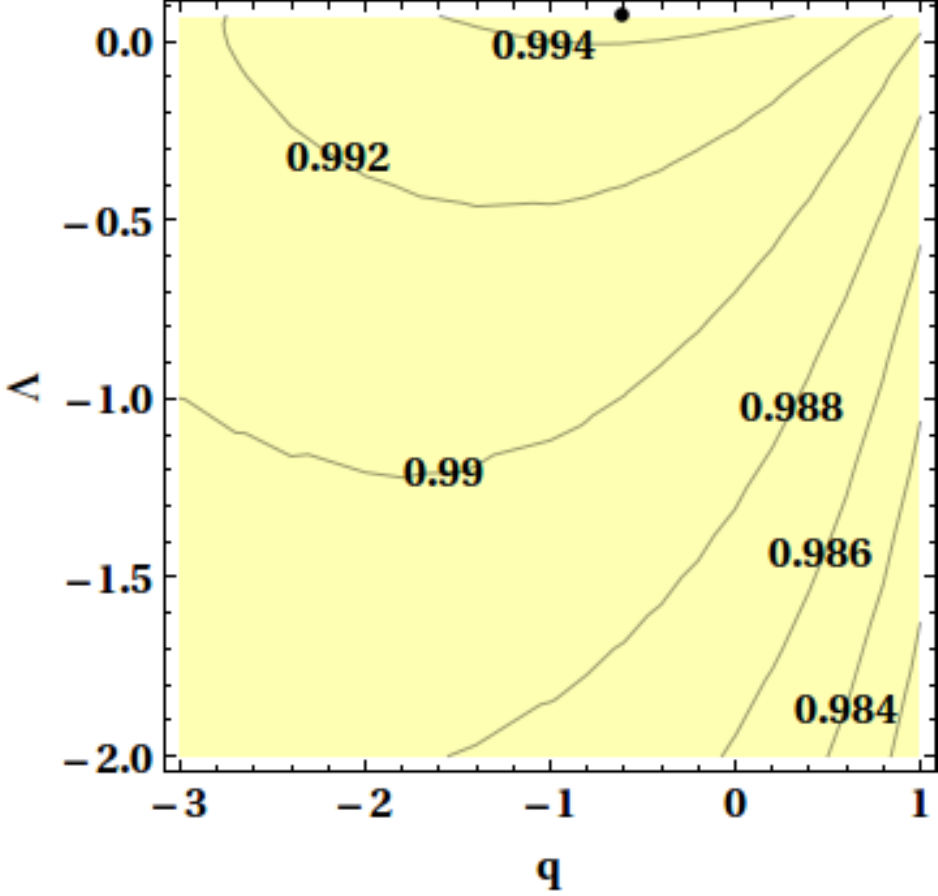}}
\hspace{3cm}
\subfloat[Modified index of agreement\label{Fig_5b}]{\includegraphics[scale=0.65]{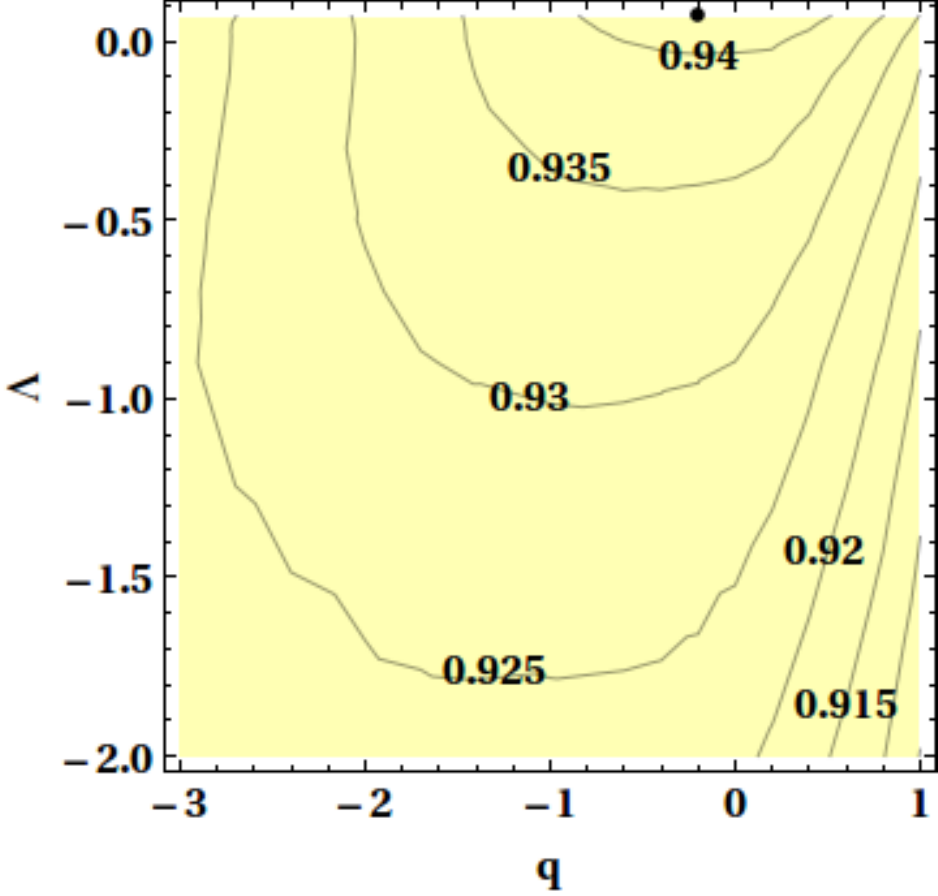}}
\caption{Figure 5: In this figure we present contours of constant (a) Index of agreement $d$ and (b) its modified form $d_1$ with variations in the tidal charge parameter $q$ and brane cosmological constant $\Lambda$. The black dot indicates the values of $q$ and $\Lambda$ where $d$ and $d_1$ attain the maximum. Both the error estimators maximize for negative values of $q$ and positive $\Lambda\sim \mathcal{O}(10^{-9})$. Note that the values of $\Lambda$ are in the units of $10^{-7} r_g^{-2}$.}
\end{figure}
It turns out that the Nash-Sutcliffe Efficiency and its modified form remains insensitive towards the differences between the observed and predicted means and variances. To overcome this shortcoming, the index of agreement is proposed \cite{willmott1984evaluation, doi:10.1080/02723646.1981.10642213,2005AdG.....5...89K,WRCR:WRCR8013}. It is denoted by $d$ and assumes the following mathematical form,
\begin{align}
d(q, \Lambda) = 1 - \frac{\sum_{i} \{ \mathcal{O}_{i} - \Omega_{i}(q, \Lambda) \}^{2}}{\sum_{i} \{ |\mathcal{O}_{i} - \mathcal{O}_{av}| + |\Omega_{i}(q, \Lambda) - \mathcal{O}_{av}| \}^{2}} \label{64}
\end{align} 
The denominator, often known as the potential error, denotes the maximum deviation of each pair of observed and predicted luminosities from the average luminosity. 

Again due to the presence of square terms in the numerator the index of agreement suffers from oversensitivity to higher values of optical luminosity and hence its  modified version $d_1$ is proposed, where,
\begin{align}
d_{1}(q, \Lambda) = 1 - \frac{\sum_{i}|\mathcal{O}_{i} - \Omega_{i}(q, \Lambda)|}{\sum_{i} \{ |\mathcal{O}_{i} - \mathcal{O}_{av}| + |\Omega_{i}(q, \Lambda) - \mathcal{O}_{av}| \}} \label{65}
\end{align}

From \ref{64} and \ref{65} it is clear that the best model for $q$ and $\Lambda$  corresponds to the maximum value for $d$ and $d_1$ which cannot be greater than 1. Since the denominators in \ref{64} and \ref{65} are greater than \ref{62} and \ref{63} respectively, the index of agreement and its modified form always assume greater values compared to $E$ and $E_1$.
\ref{Fig_5a} and \ref{Fig_5b} illustrate the constant contours of $d$ and $d_1$ with variation in $q$ and $\Lambda$. From the coordinates of the black dot which denote the maximum of $E$ and $E_1$, it is clear that the index of agreement and its modified form also attain a maxima for a negative value of $q$ and a positive $\Lambda$. The maximum value of $d$ and $d_1$ is achieved for $q\sim -0.6$ and $q\sim -0.2$ respectively. The value of $\Lambda$ that maximizes $d$ and $d_1$ corresponds to $7\times 10^{-9}r_g^{-2}$. Since \ref{Fig_5a} and \ref{Fig_5b} replicates the trend exhibited by \ref{Fig_4a} and \ref{Fig_4b} respectively, the conclusions drawn previously remain unaltered.\\
Therefore the behavior of the error estimators indicate that a negative tidal charge parameter and a positive $\Lambda$ is favored by optical observations of quasars.

\end{itemize}

\section{Summary and conclusions}\label{S6}
In this work our chief goal was to extricate the imprints of bulk $f(R)$ gravity from the quasar continuum spectrum which are ideal astrophysical probes to explore the nature of gravitational interaction in extreme situations. Extra dimensions and higher curvature corrections are often believed to be manisfestations of the ultraviolet nature of gravity with interesting consequences in inflationary cosmology, late-time cosmic acceleration, gravitational waves and collider physics. Hence, it is instructive to investigate their impact on the electromagnetic spectrum emitted by the accretion disk around quasars which are expected to exhibit maximum curvature effects, especially near the horizon. The presence of $f(R)$ gravity in higher dimensions substantially modify the effective gravitational field equations on the brane such that they evince significant deviations from Einstein's equations. Even in the absence of any matter-energy on the brane, the electric part of the Weyl tensor which represents the non-local gravitational effects of the bulk acts a source for gravity in four dimensions. In addition, the interplay of the bulk cosmological constant, the brane tension and the higher curvature terms in the bulk action naturally induce a cosmological constant in the brane whose origin is physically motivated. A positive cosmological constant is often invoked to interpret the observations related to distant Type Ia supernovae and the anisotropies in the cosmic microwave background radiation which signify an accelerated expansion of the universe. Therefore, the effect of such a term in the black hole continuum spectrum is worth exploring. 

As a first approximation, static, spherically symmetric and vacuum solutions of these modified field equations are explored since they represent the simplest deviation from the standard Schwarzschild scenario in \gr.$~$These approximations permit a decomposition of the electric part of the Weyl tensor into ``dark radiation" and ``dark pressure", such that  
various equations of state connecting them lead to different classes of black hole solutions. We consider two such solutions in this work corresponding to equations of state $P=0$ and $2U+P=0$. While the former leads to a perturbative solution, the latter assumes an exact black hole spacetime bearing a striking resemblance with the \RN de Sitter/anti-de Sitter/flat metric in \gr. The asymptotic character of the exact solution is determined by the signature and the magnitude of the brane cosmological constant while the trademark of extra dimensions is encoded in the charge parameter which can assume a negative sign unlike GR. Although we analyze the effect of both the backgrounds on the quasar continuum spectrum we perform a comparison with observations only with the exact spacetime since the perturbative background is subject to vary with higher order corrections to the metric.

It is important to note that the exact solution is characterized by two parameters namely, the tidal charge parameter and the brane cosmological constant.  
In a previous work \cite{Banerjee:2017hzw} we explored the sole impact of the charge parameter on the continuum spectrum of eighty quasars to infer that optical observations of quasars favor a \emph{negative} charge parameter. This work is subsequently generalized to axi-symmetric spacetimes \cite{Banerjee:2019sae} where the metric resembles the familiar \KN solution in GR. Inclusion of black hole rotation not only corroborates our earlier finding but also enables us to estimate the spin of the quasars \cite{Banerjee:2019sae}. This is further supported by the study of quasi-periodic oscillations in the black hole power spectrum where a negative charge parameter is reported to be favored by observations \cite{Stuchlik:2008fy}.

The present work aims to examine the effect of the charge parameter on the continuum spectrum in presence of the brane cosmological constant. In order to accomplish this we compute the theoretical estimates of optical luminosity for the sample of eighty quasars by varying the two relevant metric parameters ($q$ and $\Lambda$) and compare them with the corresponding observed values. By computing several error estimators, namely chi-squared, Nash-Sutcliffe efficiency, index of agreement and the modified versions of the last two we conclude that optical observations of quasars indeed favor a negative tidal charge parameter and a small positive $\Lambda$. A negative charge parameter which potentially arises in a higher dimensional scenario marks a clear deviation from \gr\ and this is in accordance with our previous findings. A positive $\Lambda$ on the other hand is in concordance with the aforementioned cosmological observations. We further mention that the tidal charge is related to the compactification scale of the extra dimension since the signature and magnitude of $q$ determines the extent of the extra dimension as well as the penetration of the horizon of the black hole into the bulk spacetime \cite{Chamblin:2000ra}. This is achieved by evolving the brane metric to the bulk which involves studying the evolution of the extrinsic curvature along the extra dimension. In particular, negative values of $q$ modifies the bulk metric from a black string to a black cigar thereby making the extra dimension more compactified and hence seems to be more natural \cite{Chamblin:2000ra}. In particular, the observationally favored value of $q\sim -0.6$ makes the extra dimension compactified by 1\% compared to the general relativistic scenario. The compactification scale however, continues to be the Planck scale.

Our analysis also enables us to provide an estimate on the magnitude of $\Lambda$ from the quasar optical data, which turns out to be $\mathcal{O}(10^{-9})$ in units of inverse square of the gravitational radius $r_g$. Since $r_g$ varies with the mass of the quasar, it might appear that $\Lambda$ deduced by us is mass dependent. However, one can verify that this choice of units does not affect the order of magnitude estimate of $\Lambda$ which is based on the maximum mass $M_{max}$ of the quasar in the sample. For lower mass quasars with mass $M$ we should have ideally chosen a cosmological constant $M^2/M_{max}^2$ times smaller than the $\Lambda$ of $M_{max}$ while performing the error analysis. Since $\Lambda$ is inherently very tiny, one can check that it will be even smaller for the low mass quasars and therefore, their impact on the spectrum will be negligible. With $M_{max}\sim 10^{9} M_\odot$, it can be shown that $\Lambda \sim 10^{-38} cm^{-2}$ and it is remarkable that such a tiny value of the cosmological constant can be discerned from the accretion data. A variation of the outer radius $r_{out}$ allows us to consider a marginally higher value for the repulsive $\Lambda$ which enhances the magnitude of estimated $\Lambda$ roughly by an order. Our analysis therefore establishes a strong constraint on the upper limit of $\Lambda$ from the quasar optical data. Note that this is a much stronger constraint compared to the work of P\'{e}rez et al. \cite{Perez:2012bx}, which is based on the observation of only a single stellar-mass black hole source Cygnus X-1. With enhanced precision in observing the inner regions of the disk by future telescopes and including the effects of the corona in modelling the spectral energy distribution of the quasars, a tighter constraint on $\Lambda$ can be established. A similar analysis on a different sample of quasars and micro-quasars with known masses and accretion rates is also worth exploring.

\section*{Acknowledgements}
The research of SSG is partially supported by the Science and Engineering
Research Board-Extra Mural Research Grant No. (EMR/2017/001372), Government of India.
The authors thank Dr. Sumanta Chakraborty for useful comments and discussions throughout the course of this work.

\bibliography{accretion,torsion,Gravity_1_full,Gravity_2_full,Gravity_3_partial,Brane,KN-ED}

\bibliographystyle{./utphys1}

\end{document}